\documentclass[acmsmall]{acmart}
\AtBeginDocument{%
  }

\setcopyright{acmlicensed}
\copyrightyear{2025}
\acmYear{2025}
\acmDOI{XXXXXXX.XXXXXXX}

\acmJournal{JACM}
\acmVolume{37}
\acmNumber{4}
\acmArticle{111}
\acmMonth{8}

\usepackage{titlesec}
\titlespacing{\section}{0pt}{*0.5}{*0.5}  % Adjust space before/after section
\titlespacing{\subsection}{0pt}{*0.4}{*0.4}
\titlespacing{\subsubsection}{0pt}{*0.2}{*0.2}
\usepackage{graphicx}  % For resizing tables if needed
\usepackage{booktabs}  % For professional-looking tables
\usepackage{wrapfig}
\usepackage{amsfonts}

\usepackage{float}  % Allows forcing table position
\usepackage[colorinlistoftodos,prependcaption,textsize=small]{todonotes}

\begin{document}

\title{Software Vulnerability Analysis Across Programming Language and Program Representation Landscapes: A Survey}

\author{Zhuoyun Qian}
\email{zhuoyunqian19@gmail.com}
\author{Fangtian Zhong}
\authornote{Corresponding author}
\email{fangtian.zhong@montana.edu}
\affiliation{%
  \institution{Montana State University}
  \city{Bozeman}
  \state{MT}
  \country{USA}
}

\author{Qin Hu}
\email{qhu@gsu.edu}
\author{Yili Jiang}
\email{yjiang27@gsu.edu}
\affiliation{%
  \institution{Georgia State University}
  \city{Atlanta}
  \state{Georgia}
  \country{USA}}

\author{Jiaqi Huang}
\email{jhuang@ucmo.edu}
\affiliation{%
  \institution{University of Central Missouri}
  \city{Warrensburg}
  \state{Missouri}
  \country{USA}
}

\author{Mengfei Ren}
\email{mengfei.ren@uah.edu}
\affiliation{%
 \institution{University of Alabama}
 \city{Huntsville}
 \state{Alabama}
 \country{US}}

\author{Jiguo Yu}
\email{jiguoyu17@uestc.edu.cn}
\affiliation{%
  \institution{Universtiy of Electronic Science and Technology of China}
  \city{Chengdu}
  \state{Sichuan}
  \country{China}}

\renewcommand{\shortauthors}{Zhuoyun et al.}

\begin{abstract}
Modern software systems are developed in diverse programming languages and often harbor critical vulnerabilities that attackers can exploit to compromise security. These vulnerabilities have been actively targeted in real-world attacks, causing substantial harm to users and cyberinfrastructure. Since many of these flaws originate from the code itself, a variety of techniques have been proposed to detect and mitigate them prior to software deployment. However, a comprehensive comparative study that spans different programming languages, program representations, bug types, and analysis techniques is still lacking. As a result, the relationships among programming languages, abstraction levels, vulnerability types, and detection approaches remain fragmented, and the limitations and research gaps across the landscape are not clearly understood. This article aims to bridge that gap by systematically examining widely used programming languages, levels of program representation, categories of vulnerabilities, and mainstream detection techniques. The survey provides a detailed understanding of current practices in vulnerability discovery, highlighting their strengths, limitations, and distinguishing characteristics. Furthermore, it identifies persistent challenges and outlines promising directions for future research in the field of software security.

\end{abstract}

\begin{CCSXML}
<ccs2012>
 <concept>
  <concept_id>00000000.0000000.0000000</concept_id>
  <concept_desc>Do Not Use This Code, Generate the Correct Terms for Your Paper</concept_desc>
  <concept_significance>500</concept_significance>
 </concept>
 <concept>
  <concept_id>00000000.00000000.00000000</concept_id>
  <concept_desc>Do Not Use This Code, Generate the Correct Terms for Your Paper</concept_desc>
  <concept_significance>300</concept_significance>
 </concept>
 <concept>
  <concept_id>00000000.00000000.00000000</concept_id>
  <concept_desc>Do Not Use This Code, Generate the Correct Terms for Your Paper</concept_desc>
  <concept_significance>100</concept_significance>
 </concept>
 <concept>
  <concept_id>00000000.00000000.00000000</concept_id>
  <concept_desc>Do Not Use This Code, Generate the Correct Terms for Your Paper</concept_desc>
  <concept_significance>100</concept_significance>
 </concept>
</ccs2012>
\end{CCSXML}

\ccsdesc[500]{Software Security}
\ccsdesc[300]{System security}
\ccsdesc{Vulnerability discovery}
%\ccsdesc[100]{Programming language}

%\keywords{Software security, }

\maketitle

\section{Introduction}
\subsection{Background}
Software vulnerability discovery is the cornerstone of securing modern software systems. It plays a critical role in software security, as vulnerabilities can result in unauthorized access, data breaches, system crashes, privilege escalation, and degraded user experience. By systematically identifying weaknesses in code, configuration, and execution environments, vulnerability discovery enables organizations to assess and mitigate risks before they are exploited by adversaries. In offensive security research, this process reveals exploitable code paths and vulnerable inputs, facilitating the development of targeted mitigations and comprehensive security evaluations. In defensive contexts, it supports proactive protection by uncovering previously unknown vulnerabilities, guiding patch prioritization, and enhancing system resilience against attacks. 

Vulnerability analysis is typically conducted at three levels: \textit{binary}, \textit{intermediate representation (IR)}, and \textit{source code}, with each offering unique advantages and facing specific challenges. Binary-level analysis operates directly on compiled executables, making it indispensable for assessing closed-source software and deployed applications. Techniques such as binary instrumentation, fuzzing, and taint analysis are commonly used to detect memory corruption, missing-check vulnerabilities, and side-channel leaks. However, this level suffers from limited semantic information, obfuscation challenges, and difficulties in reconstructing accurate control and data flow. IR-level analysis is performed on representations like low-level virtual machine (LLVM) IR, bytecode, assembly, and abstract syntax trees (ASTs), striking a balance between low-level execution details and high-level code semantics. Techniques such as flow-sensitive analysis, data-flow analysis, and symbolic execution are employed to detect concurrency bugs, missing-check vulnerabilities, and memory safety issues. While IR-level analysis benefits from platform independence and better analyzability than binaries, it may lack complete context in the source code. Source-code level analysis focuses on human-readable source code, allowing early detection of logic errors, insecure coding patterns, and API misuse. Common techniques include static analysis and machine learning-based vulnerability detection. This level provides rich syntactic information, enabling precise reasoning about code behavior. However, it may be limited in capturing vulnerabilities that manifest only during execution or after compiler transformations. Furthermore, the programming language itself significantly influences the nature of vulnerabilities and the detection strategies employed. For example, C/C++ programs are particularly prone to memory safety issues, performance bottlenecks, concurrency bugs, and logic flaws. In contrast, JavaScript frequently exhibits logic bugs, missing-check vulnerabilities, and concurrency-related issues due to its dynamic and event-driven nature. As such, a comprehensive understanding of software vulnerabilities requires cross-language, cross-representation, and multi-bug-type perspectives.

\subsection{Classification by Bugs}

		\begin{figure*}[h]
		\centering
		\includegraphics[scale=0.37]{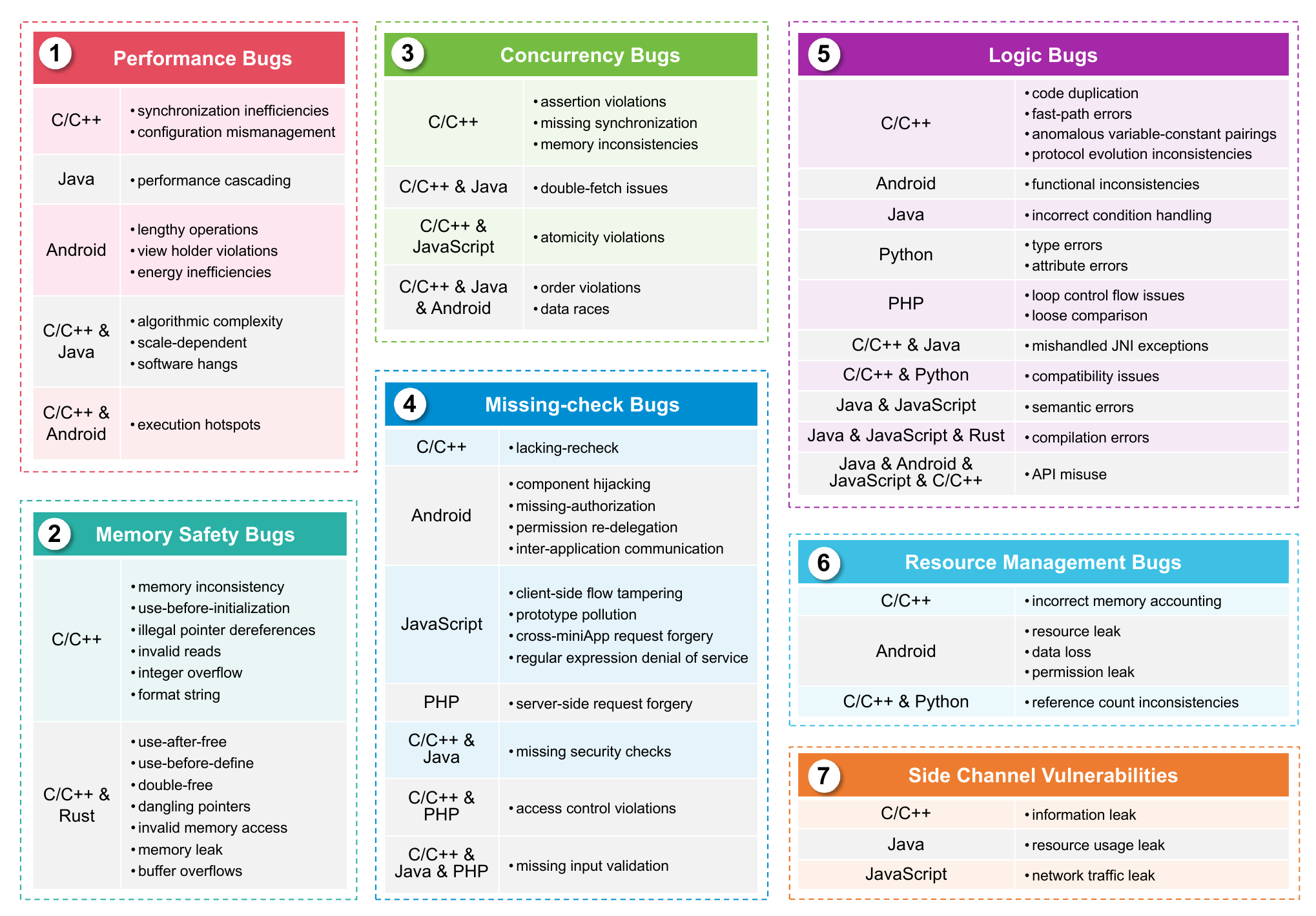}
		\caption{Classification of Bugs}
		\label{fig:bug}
	\end{figure*}
    
The classification of software bugs, as illustrated in Fig. \ref{fig:bug}, spans seven major categories: performance bugs, memory safety bugs, concurrency bugs, missing-check bugs, logic bugs, resource management bugs, and side channel vulnerabilities, each representing distinct causes of system instability or inefficiency. Performance bugs occur when software exhibits slow execution or excessive resource use without functional errors, often due to synchronization issues, configuration problems, or inefficient algorithms, frequently observed in C/C++, Java, and Android. Memory safety bugs arise from improper memory operations like use-after-free, buffer overflows, and invalid access, which may lead to crashes or data corruption, especially in C/C++ and Rust. Concurrency bugs result from incorrect handling of multiple threads or processes, leading to race conditions, atomicity violations, or inconsistent states—common across C/C++, Java, Android, and JavaScript. Missing-check bugs stem from the failure to validate inputs, permissions, or security conditions, enabling vulnerabilities such as access control violations and component hijacking across C/C++, Java, Android, PHP, and JavaScript. Logic bugs are caused by flawed control or computation logic, including API misuse, semantic errors, and improper comparisons, impacting correctness in languages like C/C++, Java, JavaScript, Python, and PHP. Resource management bugs involve the incorrect handling or accounting of system resources (e.g., memory), leading to leaks or data loss in Android, C/C++, and Python. Side-channel vulnerabilities are security flaws where attackers infer sensitive information through indirect behaviors such as timing, cache access, or traffic patterns, with examples like SSL/TLS leaks in C/C++, resource usage leaks in Java, and network traffic leaks in JavaScript.

\subsection{Related Work}

Prior surveys have explored specific aspects of software vulnerability detection, often focusing on narrow domains or limited bug types. For instance, \cite{johari2012survey} evaluates static and dynamic analysis techniques for detecting SQL injection (SQLi) and cross-site scripting (XSS) vulnerabilities, while \cite{ullah2020ss7} integrates rule-based filtering with machine learning to detect attacks in SS7 telecommunication networks. Fuzzing has been extensively studied in \cite{zhao2024systematic}, which categorizes modern fuzzing strategies, including machine learning-assisted fuzzing, constraint-solving integration, and directed fuzzing for improving code coverage and vulnerability discovery efficiency. \cite{ji2018coming} provides an overview of automated techniques for vulnerability detection, exploitation, and patching, emphasizing the growing role of machine learning in software security automation. Similarly, \cite{zuo2024vulnerability} investigates the mining of patch commits to identify silent vulnerabilities in open-source software, leveraging NLP and graph-based machine learning approaches. Machine learning has indeed become a transformative force in vulnerability detection. \cite{zhu2023application,jia2021machine} surveys deep learning-based techniques, analyzing models such as CNNs, RNNs, and Transformers for software vulnerability detection. \cite{kaur2023detection} focuses specifically on LSTM-based models for XSS detection. In contrast to these focused studies without generalizability and reusability of findings, our survey provides a broad, technique-driven overview of software vulnerabilities across multiple programming languages, program representations, and bug types. This holistic perspective fills an important gap in the current literature by integrating language diversity, representation levels, and vulnerability taxonomies into a single cohesive framework. The differences between existing surveys and ours are presented in Table \ref{tab:survey}.

\begin{table*}[h]
    \centering
    \caption{Comparison of Surveys}
    \vspace{-0.5em}
    \resizebox{\linewidth}{!}{  
    \begin{tabular}{|c|c|c|c|c|}
        \hline
        \textbf{Paper} & \textbf{Multi-Programming Languages} & \textbf{Multi-Program Representations}&\textbf{Multi-Bug Types}& \textbf{Multi-Analysis Methods}\\
       \hline
          \cite{johari2012survey}& \checkmark  & $\times$ & $\times$ & \checkmark\\
        \hline
          \cite{ullah2020ss7} & $\times$  & $\times$ & $\times$ & $\times$ \\
        \hline
 \cite{zhao2024systematic}& $\times$  & $\times$  & \checkmark & $\times$ \\
        \hline
       \cite{ji2018coming}& $\times$  & $\times$  & \checkmark & \checkmark \\
        \hline
         \cite{zuo2024vulnerability}& $\times$  & $\times$  & \checkmark & \checkmark\\
        \hline
       \cite{zhu2023application}& $\times$  & $\times$ & \checkmark & $\times$\\
       \hline
      \cite{kaur2023detection}&  $\times$ & $\times$ & $\times$ & $\times$ \\
       \hline
        Ours&  \checkmark & \checkmark& \checkmark& \checkmark \\
        \hline
    \end{tabular}
    }
    \label{tab:survey}
\end{table*}

\subsection{Contribution of the Survey}
%We differentiate the methods applied to binary, IR and source code by checking if dynamic analysis is dominated and native code is executed (this is binary). Dynamic analysis is dominated but bytecode is executed (IR) and static analysis is dominated and performed on IR. If dynamic analysis first, then static analysis (different). 
	\begin{figure*}[h]
		\centering
		\includegraphics[scale=0.45]{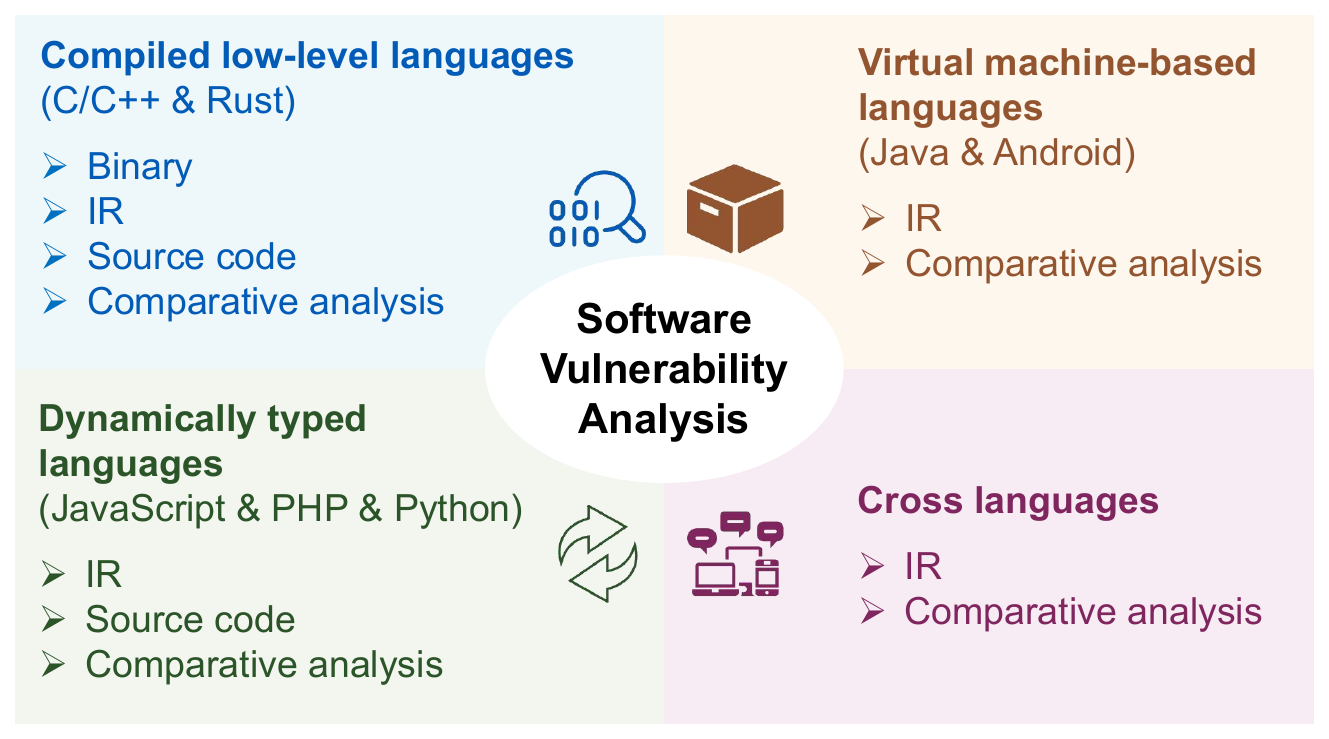}
		\caption{Overview of The Survey}
		\label{fig:overview}
	\end{figure*}
    
As shown in Fig. \ref{fig:overview}, our survey identifies the eight most vulnerability-prone programming languages, categorizing them into four groups: compiled low-level languages, virtual machine-based languages, dynamically typed languages, and cross-languages. The key contributions of this survey are outlined as follows:

\begin{enumerate}
  \item 
    \par We provide a modern, technique-driven overview of software vulnerabilities across various programming languages, program representations, and bug types, with a strong focus on detection methods, including static, dynamic, hybrid, and machine learning approaches. 
  \item 
    \par We summarize the causes of vulnerabilities and categorize them into major bug types, offering a broad and unified view of existing software flaws. Our survey highlights practical detection techniques to each bug type and introduces a language-aware categorization framework.

  \item 
    \par We provide an in-depth discussion of both well-established and underexplored research directions in this domain. We conclude by identifying future opportunities in advancing  vulnerability detection techniques.

  \item 
    \par We automate the retrieval of relevant publications on software vulnerabilities from major academic databases, including IEEE, ACM, and Google Scholar, to ensure the comprehensiveness and reproducibility of our survey. For each reviewed paper, we verify the availability of datasets and code, categorize the datasets into private, partially public, and public, and compile both the collected metadata and our automation scripts in a public repository: \url{https://github.com/fangtian-zhong/SoftwareSecurity.git}.

\end{enumerate}

\subsection{Paper Organization}
The remainder of this article is organized as follows. Sections \ref{sec:cbinary}–\ref{sec:csource} discuss bug detection techniques at the binary, IR, and source code levels along with a comparative analysis of static, dynamic, hybrid, and machine learning-based approaches for compiled, low-level languages such as C/C++ and Rust. Section \ref{sec:javair} focuses on bug detection at the IR level for virtual machine-based languages, specifically Java and Android, including a corresponding comparison of detection methodologies. Sections \ref{sec:jsir}–\ref{sec:jssource} examine bug detection at both the IR and source code levels for dynamically typed languages (\emph{i.e.}, JavaScript, PHP and Python), with comparative insights into their detection techniques. Section \ref{sec:crossir} explores bug detection at the IR level across multiple languages, again providing a comparison of different analysis approaches. Finally, Section \ref{sec:trends} presents key challenges and open research gaps, followed by the conclusion in Section \ref{sec:conclusion}.

\section{C/C++ \& Rust Bug Detection at Binary Level} \label{sec:cbinary}

\subsection{Logic bugs} Logic bugs at the binary level include compatibility issues and semantic errors, detected through static and dynamic analysis. \textbf{Static analysis} identifies compatibility issues in external libraries via pattern matching, flagging modified interfaces while surrounding code remains unchanged \cite{jia2021depowl}. It uses ABI-Tracker to detect backward and forward application binary interface (ABI) changes, such as added, removed, or altered interfaces. If an application relies on an incompatible library version, it is flagged as a compatibility issue. \textbf{Dynamic analysis} uses fuzzing to detect semantic inconsistencies in software verification tools, including code coverage tools \cite{yang2019hunting}, SMT solvers \cite{mansur2020detecting}, model checkers \cite{zhang2019finding}, compilers \cite{guo2022detecting}, and type checkers \cite{dewey2015fuzzing}. For code coverage tools like gcov and llvm-cov, fuzzing with Csmith generates random C programs, detecting discrepancies in execution counts and line coverage. In SMT solvers, fuzzing breaks seed formulas into sub-formulas, assigns random truth values, mutates them, and checks if solvers misclassify satisfiable instances. For C model checkers, fuzzing mutates seed programs, inserts counter variables, and generates program-property pairs to detect inconsistencies. In Simulink compilers, fuzzing applies assertion injection and block mutation, selectively mutating unexecuted blocks using MCMC sampling while maintaining functional equivalence. Discrepancies between original and mutated outputs indicate compiler errors. For type checkers, fuzzing generates well-typed programs using constraint logic programming and type rule encoding, then mutates them to analyze checker behavior. It detects precision bugs (valid programs rejected), soundness issues (ill-typed programs accepted), and consistency bugs (conflicting results). Fuzzing, combined with taint analysis and field-sensitive data-flow analysis, detects configuration-related logic errors by identifying the internal and external effects\cite{wang2023understanding}. It dynamically mutates configuration options, executes the program under startup-loaded and modified configurations, and compares their output to assess external effect. To track internal effects, taint analysis and field-sensitive data-flow analysis monitor global configuration variable values.

\subsection{Missing-check bugs}
Missing-check bugs at the binary level include missing security checks, detected through hybrid analysis. \textbf{Hybrid analysis:} It integrates intra-procedural data-flow analysis with fuzzing to detect access control violations \cite{hu2021achyb}. Kernel source code is compiled into LLVM IR, constructing DFGs and performing interface analysis to identify permission checks. Intra-procedural data-flow analysis tracks privileged function calls, while constraint-based invariant analysis verifies permission conditions. Greybox fuzzing with distilled seeds is applied, using a clustering-based seed distillation technique to prioritize test cases likely to expose vulnerabilities. 

\subsection{Concurrency bugs} Concurrency bugs at the binary level include order violations, data races, atomicity violations, missing synchronization, memory inconsistencies, double-fetch issues, and assertion violations, detected through dynamic and hybrid analysis. \textbf{Dynamic Analysis:}  \textit{Execution monitoring} detects concurrency-induced order violations by recording thread interactions and tracking memory operations, identifying null pointer dereferences (NPD), double pointers (DP), uninitialized reads, and buffer overflows \cite{zhang2013conmem}. It validates buggy interleavings by instrumenting and replaying them while also detecting non-interleaved concurrency bugs through different execution models. AsyncAsSync replaces asynchronous calls with normal procedure calls to simulate sequential execution, while AsyncAsEvents randomizes asynchronous execution order. AsyncGeneral combines both methods, first detecting non-interleaved concurrency bugs by collecting code traces from AsyncAsSync and comparing them with the original program, and then uncovering synchronization errors with AsyncGeneral \cite{joshi2012underspecified}. \textit{Controlled
execution perturbation} detects exchangeable memory operations by forcing execution scheduling to identify UAF, NPD, and DF bugs. It checks if a free(p) operation and all accesses to p are exchangeable, profiles memory access traces to detect unserializable interleavings, and confirms atomicity violations by injecting delays to force rare interleavings leading to crashes or incorrect outputs \cite{cai2019detecting,yu2021detecting,park2009ctrigger}. Sequential tests are converted into concurrent client requests to explore interleavings, detecting bugs through output discrepancies \cite{fonseca2011finding}. \textit{Fuzzing} generates high-coverage inputs and guides execution toward risky interleavings. It instruments memory and persistency instructions with LLVM and applies an AFL++-based mutator to prioritize read-after-write interleavings, exposing non-persisted data reads and durable side effects while monitoring synchronization variables for inconsistencies. Memory aliasing and branch coverage refine interleaving exploration\cite{chen2022efficiently}.  Fuzzing also generates concurrent tests ensuring writer-reader pairs access the same memory, exposing data races, order violations, and atomicity violations \cite{gong2021snowboard}. It can be combined with side-channel analysis to detect double-fetch bugs, where memory is fetched multiple times without synchronization, leading to race conditions \cite{schwarz2018automated}.  Flush+Reload flushes memory locations from the CPU cache and measures reload times to detect multiple fetches. Syscall fuzzing identifies privilege escalation cases when memory is modified between fetches. \textbf{Hybrid analysis:} Hybrid techniques integrate data-flow analysis, static slicing, fuzzing, execution monitoring, record-and-replay, and controlled execution perturbation. One approach combines \textit{fuzzing} with \textit{data-flow analysis} to detect concurrency relationships triggering buffer overflows, DF, or UAF errors \cite{liu2018heuristic}. It constructs data-flow graphs to track shared variable accesses across threads and manipulates execution order to expose concurrency-induced errors by fuzzing. Another approach integrates \textit{static slicing} with \textit{execution-trace analysis} to track shared-memory reads contributing to infinite loops, assertion violations, and memory errors \cite{zhang2011conseq}. It traces write operations one step back in the data dependency chain from read instructions and replays suspicious interleavings to confirm concurrency bugs. \textit{Record-and-replay} is combined with \textit{controlled execution perturbation} to detect data races  \cite{yuan2021raproducer}. Static analysis identifies locks based on intrinsic characteristics, and execution traces are replayed with a proof-of-concept (PoC) input. Shared variables accessed during execution are analyzed, and well-synchronized accesses are eliminated based on detected locks. Race conditions are then actively forced through thread scheduling perturbation.

\subsection{Memory Safety Bugs} Memory safety bugs at the binary level include memory inconsistency, buffer-overflow, use-after-free (UAF) , use-before-define (UBD),  double-free (DF), dangling pointers (DP), and invalid memory access errors, detected through dynamic analysis, hybrid analysis, and machine learning. \textbf{Dynamic analysis:} \textit{Failure injection} introduces controlled faults to evaluate memory state and recoverability \cite{liu2020cross,gonccalves2023mumak}. It tracks read/write operations, logging memory addresses before and after failure to detect memory inconsistencies. A failure point tree maps faults to call stacks, injecting faults at unvisited points and re-executing the program to test recovery. \textit{Pattern matching} detects memory inconsistencies by tracking memory stores, cache flushes, and fences, mapping memory locations to operations and applying predefined rules (e.g., missing durability guarantees, multiple overwrites, ordering violations) \cite{di2021fast}.  \textit{Fuzzing}  generates test cases and apply backward taint analysis to track memory accesses \cite{wang2019locating}. Buffer overflows are identified by reconstructing valid memory layouts from successful executions and detecting boundary violations in failed executions, highlighting discrepancies between memory states. It detects UAF, UBD, DF, DP, and invalid memory access errors in library APIs by generating test cases that trigger undefined behaviors \cite{takashima2021syrust,takashima2024crabtree}. It synthesizes valid test cases based on API typing specifications (input parameters, return types), maps raw byte data to API parameters, and inserts API call sequences. API specifications are refined by adjusting polymorphic function signatures to mitigate recurring compiler errors, and a coverage- and type-guided search algorithm optimizes API sequence generation. The test cases are executed with generated inputs, and undecided behaviors are flagged as bugs. \textbf{Hybrid analysis} This approach combines typestate, data flow analysis and taint analysis with fuzzing, runtime execution and record-and-replay to detect UAF and UBD bugs \cite{wang2020typestate,chen2017pinpointing}. Typestate-guided \textit{fuzzing} constructs operation sequences such as $\langle \text{malloc} \rightarrow \text{free} \rightarrow \text{use} \rangle$ from \textit{typestate analysis}, instruments the programs, and adaptively mutates inputs using flow-sensitive information to maximize sequences coverage. \textit{Static taint analysis}  applies track integer operations affecting memory allocations on GIMPLE IR , inserts overflow checks and guarding code at memory sites \cite{sun2015efficient}. Memory allocation behavior is monitored at runtime to flag inputs causing overflows. \textit{Record-and-replay} with \textit{data-flow analysis} logs system calls and execution traces, identifying anomalies via crash detection and syscall interposition. Anomalous traces undergo dynamic binary translation, instrumenting the program to log runtime memory access. The instrumented program is replayed to extract data flow and compare it with a precomputed data flow graph, detecting uninitialized memory access. \textbf{Machine learning:} \textit{Deep reinforcement learning} enhances \textit{fuzzing} to detect buffer overflows via \textit{concolic execution} \cite{jeon2022dr}. Coverage-guided fuzzing explores programs, marking test cases that reach new basic blocks. These cases are prioritized by reinforcement learning with a Deep Q-network for unexplored path execution, with constraints solved using an SMT solver to guide fuzzing until a crash occurs.

\subsection{Performance Bugs} Performance bugs at the binary level include synchronization, algorithm complexity, configuration, scale-dependent, and timeout-related issues, detected through dynamic analysis and machine learning. \textbf{Dynamic analysis:} \textit{Synchronization analysis} detects inefficiencies in multithreaded applications by intercepting pthread synchronization primitives at runtime, tracking lock usage and contention points to identify improper primitives, granularity issues, over-synchronization, asymmetric contention, and load imbalance \cite{alam2017syncperf}. Resource-usage-guided \textit{fuzzing} monitors functions, basic blocks, and edge execution in the control flow graph while tracking resource usage \cite{petsios2017slowfuzz}. It mutates inputs using random and dictionary-based modifications, prioritizing those that consume excessive resources or cause slowdowns, exposing algorithmic complexity vulnerabilities. \textbf{Machine learning:} Configuration, scale-dependent, and timeout-related issues are detected through natural language processing (NLP), supervised learning, and unsupervised anomaly detection. \textit{NLP} extracts configuration rules from documentation and compares runtime performance against expected outcomes under different configurations to identify anomalies \cite{he2020cp}. Supervised anomaly detection trains a \textit{regression model} on control features (e.g., input properties, configuration parameters) and observational features (e.g., syscall counts, libc function calls) from small-scale, bug-free runs to predict large-scale behaviors \cite{zhou2013wukong}. Performance anomalies are detected by comparing predicted and observed system behavior at scale. Unsupervised anomaly detection applies a \textit{Self-Organizing Maps model} to learn execution patterns from timeout-related system call traces, analyzing frequencies to identify anomalies causing server hangs or slowdowns \cite{he2018tscope}.

\subsection{Side-channel Vulnerability}  
Side-channel vulnerabilities at the binary level involve information leaks, detected through dynamic analysis and machine learning. \textbf{Dynamic analysis:} \textit{Execution trace analysis} identifies unintended variations in execution \cite{xiao2017stacco,cao2019principled}.   For SSL/TLS leaks, it compares execution differences between conformant and non-conformant packet sequences. Modified ClientKeyExchange messages and CBC-encrypted data are injected, and the target program executes while constructing dynamic control-flow graphs. Vulnerabilities are reported by comparing basic block traces. For TCP leaks, dynamic analysis simulates correct and incorrect secret guesses, sending identical packet sequences in both cases. A vulnerability is flagged if responses differ, indicating behavioral discrepancies that leak sensitive information. \textbf{Machine learning:} This method detects side-channel vulnerabilities by analyzing network traffic patterns and user interactions \cite{chapman2011automated}. It simulates interactions, logs network traces, and constructs a finite-state machine where each state represents a saved DOM state, and transitions capture network behaviors.  leak quantifier applies distance metrics and the Fisher criterion to classify traces by user actions, clustering them via \textit{a multi-class classifier}. Unknown traces are assigned to the closest centroid using Hamiltonian distance. If traces for different user actions are easily distinguishable, a side-channel vulnerability is reported.

\subsection{Multiple Types of Bugs}
Different types of bugs detected simultaneously at the binary level include logic bugs, memory safety bugs, and missing-check bugs, identified through dynamic analysis, hybrid analysis, and machine learning. \textbf{Dynamic analysis:} Fuzzing and dynamic symbolic execution detect multiple bug types. Hydra \cite{kim2020findingfile} mutates OS images and syscalls by \textit{fuzzing}, applies four dedicated checkers to detect bugs, validates issues via VM-based re-execution, and minimizes syscall sequences through delta debugging to produce concise PoC reports. Cerebro \cite{li2019cerebro} optimizes seed selection by scoring inputs based on execution time, file size, edge coverage, and trace complexity, prioritizing high-impact seeds for mutation. It allocates execution cycles and mutates seeds to trigger crashes.  VulScope \cite{dai2021facilitating} collects execution traces from a reference with known vulnerabilities and a target program version, mutating inputs to explore execution detours and prioritize paths with high trace similarity to trigger crashes. Fuzzing also aids bug localization. VulnLoc \cite{shen2021localizing} tracks conditional statements in a vulnerable binary, generating a focused test suite through iterative input mutation, sensitivity map inference, and execution tracing. It computes sufficiency and necessity scores for conditional statements, ranks vulnerable locations, and maps binary instructions to source code statements using objdump. Symbolic execution explores complex paths where fuzzing struggles. BitFuzz \cite{caballero2010input} employs stitched \textit{dynamic symbolic execution}, identifying encoding functions and their inverses via \textit{trace-based dependency analysis}, bypassing difficult constraints and focusing on solvable execution path constraints. It re-stitches the output using encoding functions or their inverses, generating valid inputs to uncover vulnerabilities.

\textbf{Hybrid analysis:} it enhances fuzzing by optimizing input generation and prioritizing execution paths. One approach extracts sensitivity, complexity, depth, and rarity metrics from basic blocks via \textit{intra-procedural analysis}, assigns weight values, and injects trampolines to track control-flow edges \cite{situ2021vulnerable}. \textit{Fuzzing} allocates execution energy to rare or high-weight regions, mutating inputs to maximize code coverage and trigger crashes. Another applies \textit{path-sensitive AST analysis} and reflection-based techniques to generate semantically valid JavaScript seed inputs, mutating them to explore JavaScript engines \cite{he2021sofi}. It modifies seeds by altering constants, inserting member operations, and injecting function calls, then executes them to detect bugs. Seeds that improve coverage without crashes are added for iterative fuzzing. A final approach extracts API descriptions from ASTs and applies fuzzing to iteratively generate Python applications around APIs \cite{li2023pyrtfuzz}. It refines these applications based on CFG coverage feedback, executing them within a time-constrained budget using type-aware mutations. During execution, it monitors for unexpected behaviors, testing the Python runtime for vulnerabilities.  DSFuzz \cite{liu2023dsfuzz} uncovers deep-state bugs by constructing a state dependency graph using \textit{def-use analysis, inter-procedural data-flow analysis, and dynamic data-flow analysis} to capture direct/indirect control dependencies. It identifies deep states dependent on prior states and determines dependent input bytes for each state transition via \textit{dynamic taint tracking}. Progressive micro-directed \textit{fuzzing} modifies inputs to transition from the current state to target deep states, flagging mutated seeds causing crashes as bugs. 1dFuzz \cite{yang20231dfuzz} introduces directed differential fuzzing to reproduce 1-day vulnerabilities. It disassembles patched and unpatched binaries, generates call graphs, and identifies patch locations by comparing trailing call sequence (TCS) features. Directed differential \textit{fuzzing} then generates PoC inputs for these locations, refining fuzzing with a TCS-based distance metric to improve accuracy. RPG \cite{xu2024rpg} detects logic and memory safety bugs in Rust libraries by generating fuzz targets based on API information extracted from rustdoc. It builds an API dependency graph using static analysis, applies Pool-based Sequence Generation to prioritize unsafe code while maintaining type consistency, and synthesizes fuzz targets by linking the library with AFL. \textbf{Machine learning:} ML improves fuzzing and symbolic execution by guiding execution-path exploration, optimizing input generation, and refining mutation strategies. CAMFuzz \cite{shi2022camfuzz} trains a CNN on input seeds and CFGs to predict input bytes influencing unexplored paths and optimize \textit{fuzzing} mutation strategies. It extracts constants such as file signatures and loop counts to refine mutations, improving coverage and bug detection. SyML \cite{ruaro2021syml} enhances \textit{dynamic symbolic execution} by predicting high-risk execution paths. It extracts execution trace features (e.g., memory accesses, function complexity, system calls), trains a supervised model, and guides symbolic execution toward paths most likely to contain bugs.

\subsection{Conclusion of Bug Detection at Binary Level}\vspace{-0.8em}
\begin{table*}[h]
    \centering
    \caption{Comparison of Bug Detection Techniques for C/C++ \& Rust}
    \vspace{-0.5em}
    \resizebox{\linewidth}{!}{  % Resizes table to fit within the page width
    \begin{tabular}{|c|c|c|c|c|}
        \hline
        \textbf{Paper} & \textbf{ Techniques} & \textbf{ Bugs}&Datasets&Accuracy\&Overhead \\
        \hline
        \cite{jia2021depowl}&static &C/C++, logic &public &-\\
        \hline
        \cite{yang2019hunting},\cite{mansur2020detecting},\cite{zhang2019finding},\cite{guo2022detecting},\cite{dewey2015fuzzing},\cite{wang2023understanding}&dynamic &C/C++, logic &private, public&-,-,-,68.8\%,-,75\% \\
        \hline 
    \cite{hu2021achyb}&hybrid &C/C++, missing-check &public&28.9\% TP\\
\hline
\cite{zhang2013conmem},\cite{joshi2012underspecified},\cite{cai2019detecting},\cite{park2009ctrigger},\cite{fonseca2011finding},\cite{yu2021detecting},\cite{chen2022efficiently},\cite{gong2021snowboard},\cite{schwarz2018automated}&dynamic &C/C++, concurrency &private ,public&90\%,-,66.7\%,-,-,-,-,-,97 \%\&0.8\%,\\
        \hline \cite{liu2018heuristic},\cite{zhang2011conseq},\cite{yuan2021raproducer}&hybrid &C/C++, concurrency &private,public& 115.5 \% ,90.9\%\& 662.6\%,63.3 \%\\
        \hline
        \cite{liu2020cross},\cite{wang2019locating}, \cite{gonccalves2023mumak},\cite{di2021fast}&dynamic &C/C++, memory safety &public &-, 96\%, 117.85\%,30.6\% \\
        \hline
        \cite{wang2020typestate},\cite{sun2015efficient},\cite{chen2017pinpointing}&hybrid &C/C++, memory safety &private, partially public &-,99.99\%\&0.69\%, 0.35\% \\
        \hline
        \cite{jeon2022dr}&ML&C/C++, memory safety &public&-\\
        \hline
          \cite{takashima2021syrust},\cite{takashima2024crabtree}&dynamic &Rust, memory safety &private, public &-,-\\
        \hline
        \cite{alam2017syncperf},\cite{petsios2017slowfuzz}& dynamic  &C/C++, performance &private, public   &2.3 \%,-\\
        \hline
        \cite{he2020cp},\cite{zhou2013wukong},\cite{he2018tscope}&ML  &C/C++, performance  & private, public   &70.5\%, 92.5\%\&11.4\%,94.7\%\&1\%\\
        \hline
  \cite{xiao2017stacco},\cite{cao2019principled}&dynamic &C/C++, side channel &public &-,-\\
       \hline
   \cite{chapman2011automated}&ML&C/C++, side channel &private &61.2\% \\
       \hline
    \cite{kim2020findingfile},\cite{li2019cerebro},\cite{dai2021facilitating},\cite{shen2021localizing},\cite{caballero2010input}&dynamic &C/C++, logic, memory safety &private, public&79.6\%,-,-,88.4\%,-\\
       \hline
 \cite{situ2021vulnerable},\cite{he2021sofi}&hyrbid &C/C++, logic, memory safety &partially public,public &-,-\\
       \hline
        \cite{li2023pyrtfuzz}&hyrbid &C/C++, logic, memory safety, performance &public &-\\
       \hline
        \cite{xu2024rpg}&hyrbid &Rust, logic, memory safety &public &-\\
       \hline
        \cite{shi2022camfuzz}&ML&C/C++, unlisted&private &82\% \\
       \hline
        \cite{ruaro2021syml}&ML&C/C++, logic, memory safety &public & 78.6\% \\
          \hline
\cite{liu2023dsfuzz},\cite{yang20231dfuzz}&hybrid &C/C++,unlisted &private,public &-,-\\
       \hline
    \end{tabular}
    }
    \label{tab:cbinary}
\end{table*}

As shown in Table \ref{tab:cbinary}, static analysis is less common and only remains effective for detecting logic bugs such as ABI incompatibilities. Dynamic analysis is the dominant approach for binary-level bug detection, widely applied across logic, concurrency, memory safety, performance, and side-channel vulnerabilities. Tools like Csmith-based fuzzing \cite{yang2019hunting}, \cite{zhang2019finding}, and symbolic tracing tools \cite{wang2023understanding} detect semantic inconsistencies in verification tools, SMT solvers, and compilers with up to 75\% precision. For concurrency bugs, execution perturbation and interleaving-aware fuzzing (e.g.,  \cite{gong2021snowboard}) reveal data races, atomicity violations, and synchronization errors, with tools like \cite{schwarz2018automated} achieving 97\% accuracy using cache-based side-channel monitoring. Memory safety bugs, including UAF, DF, and buffer overflows, are tackled using fuzzing, fault injection, and taint tracking. For instance, typestate-guided fuzzing tools \cite{wang2020typestate}, \cite{chen2017pinpointing} report up to 99.99\% accuracy. Rust-specific tools such as  \cite{takashima2021syrust} and  \cite{takashima2024crabtree} synthesize API sequences to expose undefined behaviors in libraries. Hybrid analysis combines static slicing, execution monitoring, and data-flow tracking to improve precision and bug coverage. Notable tools include \cite{yuan2021raproducer} for concurrency bug reproduction and  \cite{liu2023dsfuzz} for detecting deep-state multi-type vulnerabilities. ML-based approaches are increasingly used to enhance fuzzing and symbolic execution. DRL-based methods like DR \cite{jeon2022dr} improve path exploration and crash detection, while \cite{shi2022camfuzz} and  \cite{ruaro2021syml} guide fuzzing using control-flow features and execution path risk prediction, achieving up to 82\% accuracy.

\section{C/C++ \& Rust Bug Detection at IR Level}\label{sec:cir}
\subsection{Logic Bugs} 
Logic bugs at the IR level include code duplication and semantic bugs, detected through static analysis. \textbf{Static analysis:} Buggy code clones are identified by extracting ASTs, transforming them into n-dimensional characteristic vectors, and clustering similar fragments using \textit{locality-sensitive hashing} \cite{gabel2010scalable}. Structural differences between clones are analyzed to detect syntactic inconsistencies introduced during modifications. Recurring semantic bugs are identified by constructing data dependency graphs from vulnerable traces and applying \textit{taint analysis} to extract feature vectors encoding low-level properties (e.g., operator usage frequencies) and high-level behavioral signatures (e.g., buffer overflow conditions) \cite{kang2022tracer}. Vulnerability signatures are stored in a database, and new traces are flagged as potential vulnerabilities if their cosine similarity score exceeds 0.85.

\subsection{Missing-check Bugs}
Missing-check bugs detected at the IR level include missing security checks, lacking-recheck, missing input validation, detected through static and machine learning. \textbf{Static analysis:} This method constructs CGs, CFGs, and DFGs and applies data-flow, intra/inter-procedural, flow-sensitive,  field-sensitive, taint, and alias analysis. One approach detects missing security checks by compiling kernel source code into LLVM IR, generating CGs, CFGs, and DFGs, and identifying security checks through conditional branches leading to error-handling statements \cite{wang2021amchex}. A bidirectional \textit{intra/inter-procedural data-flow analysis} tracks security-sensitive operations to ensure security enforcement on all execution paths. Another approach detects lacking-recheck bugs by constructing CGs, CFGs, and DFGs from LLVM IR  and and applying error-code inference to identify security checks \cite{wang2018check}.
CFG traversal tracks critical variables using \textit{taint tracking} and \textit{alias analysis} to form check-use chains, detecting modifications after the initial check via forward \textit{interprocedural data-flow analysis}. \textbf{Machine learning:} this method integrates with taint analysis to analyze code dependencies and  behaviors \cite{yamaguchi2013chucky}. ASTs are extracted to identify sources, sinks, conditions, assignments, and API symbols. A bag-of-words model clusters functions with similar API symbols, while \textit{taint analysis} tracks data flow from security-sensitive sources using dependency graphs. Normalized tainted conditions are embedded into vector representations, and a \textit{normality model}, trained on correctly validated functions, detects anomalies by measuring deviation from expected patterns. Functions with high anomaly scores are flagged as missing input validation checks.

\subsection{Concurrency Bugs} 
Concurrency bugs at the IR level include data race, order violations, synchronization violations, and atomicity issues, detected through static and hybrid analysis. \textbf{Static analysis:} It constructs CGs, CFGs, and DFGs and applies flow-sensitive, inter-procedural, alias, and data-flow analysis. One approach detects data races by converting source code into Boogie IR, constructing control flow graphs, and applying \textit{inter-procedural analysis} to identify available locks \cite{deligiannis2015fast}. \textit{Symbolic verification} analyzes concurrent entry points, tracking locksets on shared memory using \textit{data-flow analysis}. A data race is flagged if two locksets do not intersect.
 Another approach detects sleep-in-atomic-context bugs by compiling the Linux kernel into LLVM IR, performing connection-based \textit{alias analysis} to construct a full kernel call graph, and applying summary-based inter-procedural analysis to track atomic code paths leading to sleep function calls \cite{bai2020effective}. Path-checking is used to reduce false positives. Other approaches enforce synchronization constraints and memory access ordering.  \textit{Synchronization constraints} ensure correct interrupt-related synchronization rules using (Precondition, Postcondition) annotations \cite{tan2011acomment}. It extracts interrupt-related statements from comments, assertions, and macros, propagates annotations across the call graph, and verifies whether root functions have conflicting preconditions and postconditions. Memory access ordering ensures correct execution order using \textit{memory barriers} \cite{lepers2023ofence}. It extracts barriers from source code, constructs control-flow graphs, and classifies them into unpaired barriers, paired barriers for ordering violations, and multi-paired barriers for correct multi-threaded execution. Violations are flagged for misplaced memory accesses, incorrect barrier types, and redundant reads. \textbf{Hybrid analysis:} It collects execution traces and applies flow-sensitive and alias analysis to detect concurrency bugs. One approach tracks \textit{pointer aliasing} across execution paths and mutates execution order to predict concurrency violations \cite{guo2024reorder}. Programs are instrumented to collect event traces and LLVM IR slices, mapping each event to its corresponding instruction. Pointer flows are analyzed to detect conflicting event pairs, and sequential orderings are mutated to identify NPD, UPU, UAF, and DF due to order violations.
Another approach integrates \textit{flow analysis} with execution traces, collecting control flow traces from client programs and performing core analysis on the server when crashes or deadlocks occur \cite{kasikci2017lazy}. It applies \textit{points-to analysis}, type-based ranking, bug pattern classification, and statistical diagnosis to detect order violations and atomicity issues. Failure traces are compared with successful execution traces near failure points.

\subsection{Memory safety bugs} 
Memory safety bugs at the IR level include use-before-initialization (UBI), memory leaks, illegal pointer dereferences, buffer overflows, invalid reads, UAF, integer overflow and format string vulnerabilities, detected through static, dynamic, hybrid analysis, and machine learning. \textbf{Static analysis:} It verifies memory integrity by converting code into IR, constructing CGs,  CFGs and DFGs, and applying flow-sensitive, field-sensitive, alias, intra/inter-procedural, taint and data-flow analysis along with symbolic execution. UbiTect \cite{UBITect} constructs CGs, CFGs, and DFGs from LLVM IR and performs \textit{flow- and field-sensitive intra-procedural points-to and alias analysis} to track variables and memory objects. It propagates function summaries across call sites and uses a type qualifier inference system to verify If  variables or memory object fields in the caller functions exhibit weaker-than-expected qualifiers and employs under-constrained \textit{symbolic execution} to validate  UBI bugs. NDI \cite{zhou2022non} extracts security-critical variables using \textit{intra-procedural field-sensitive data-flow analysis}, collects inconsistent path pairs leading to security state inconsistencies, and applies under-constrained \textit{symbolic execution} to detect discrepancies at merging points. \textit{Inter-procedural, flow-insensitive, and field-insensitive} program slicing at the function level refine the analysis scope along the paths from the merging point to critical variable usage. It then performs \textit{inter-procedural analysis} to determines whether inconsistencies persist. Cod \cite{guo2024precise} resolves heap aliasing issues and distinguishes symbolic heap locations to detect memory corruption. It constructs CGs and CFGs, applies \textit{intra-procedural analysis} to initialize function states, and generates function summaries capturing memory operations. \textit{Inter-procedural analysis} propagates memory effects across functions, inlining summaries when disjointness assumptions fail. \textit{SMT solving} identifies memory safety violations. Rupair \cite{hua2021rupair} detects unchecked arithmetic, unsafe memory access patterns, and lossy integer casting by converting source code into ASTs and MIR to collect information on risky operations such as raw pointer dereferences and direct memory manipulations. Using \textit{intra-procedural data-flow and alias analysis}, Rupair identifies potential buffer overflow candidates by tracking live variables in unsafe blocks and checking their definition sites for matches with overflow patterns. It translates candidate overflows into symbolic constraints and uses \textit{SMT solving} to generate concrete counterexamples to confirm buffer overflow occurrences. SafeDrop \cite{cui2023safedrop} applies \textit{taint analysis}, compiling programs into MIR, constructing CGs, CFGs, and DFGs, and simplifying CFGs using the Tarjan algorithm to prioritize bug-prone paths. \textit{Flow- and field-sensitive alias analysis} tracks composite-type variables, while \textit{inter-procedural analysis} remains context-insensitive for efficiency. It caches alias sets for function arguments and return values, scans for unsafe patterns, and propagates taint through alias sets to detect memory corruption.

 \textbf{Dynamic analysis:} it detects memory safety bugs through execution tracing. Safe Sulong \cite{rigger2018sulong} interprets LLVM IR in Java to detect out-of-bounds accesses, UAF errors, and invalid memory deallocation in C programs. It compiles source code into LLVM IR using Clang, interprets it with Truffle to generate execution traces, and compiles frequently executed ASTs to machine code. Execution is terminated upon detecting a memory violation, and an error is reported. \textbf{Hybrid analysis:} it integrates symbolic execution with execution tracing to verify memory integrity. Bunkerbuster \cite{yagemann2021automated} reconstructs symbolic program state traces from PT-enabled kernel drivers, filters potentially buggy traces, and applies \textit{symbolic execution} on VEX IR to symbolize memory snapshots. By analyzing executed basic blocks, it detects buffer overflows, UAF errors, and format string vulnerabilities. Symbolic root cause analysis then identifies the precise origin of memory safety violations. \textbf{Machine learning:} It enhances inter-procedural analysis for detecting memory-related bugs at the statement level \cite{cao2022mvd}. It extracts Program Dependence Graphs (PDGs) from source code, incorporating control-flow and data-flow graphs to capture execution dependencies. PDGs are extended with call relations and return values from call graphs for \textit{inter-procedural analysis}. Backward and forward program slicing from key points, such as system API calls and pointer variables, generates focused program slices. Each statement node is vectorized using Doc2Vec, and a flow-sensitive graph neural network model is trained to identify bug patterns from these graph-based representations.

\subsection{Resource Management Bugs} Resource management bugs at the IR level include inconsistent reference count and memory accounting issues. \textbf{Static analysis:} One approach employs inter-procedural analysis, field-sensitive analysis, and symbolic execution to detect inconsistent reference count bugs \cite{mao2016rid}. It ensures reference count values remain non-negative and eventually reach zero by compiling source code into Linux bitcode, constructing CGs and CFGs, and applying \textit{summary-based inter-procedural analysis} to track modifications. The \textit{Z3 solver} enforces constraints on function paths, flagging bugs when execution paths with identical constraints show conflicting reference count updates. Another approach detects memory accounting bugs caused by improper tracking or deallocation of allocated memory, leading to system memory counter imbalances \cite{yang2022making}.  It compiles the Linux kernel into LLVM bitcode, constructs a kernel call graph, and identifies memory accounting interfaces by matching counter variables within kernel IR instructions. \textit{Data-flow analysis} maps allocation and free operations to corresponding accounting structures, while \textit{inter-procedural bitwise data-flow tracing} analyzes accounting flag conditions to handle conditional memory tracking. Bugs are identified by detecting failures in updating system counters during allocation or deallocation.

\subsection{Multiple Types of Bugs} 
Different types of bugs detected at the IR level include logic bugs, memory safety bugs, missing-check bugs, resource management bugs and side-channel vulnerabilities. Detection techniques rely on static analysis, and machine learning. \textbf{Static analysis:} it detects various bugs by constructing CFGs and DFGs and applying taint analysis, data-flow analysis, and signature matching. One approach applies \textit{signature matching} to detect unpatched binaries \cite{xu2020patch}, disassembling binaries into assembly code, generating CFGs, and comparing vulnerable and patched binaries to extract patch signatures. If a function in the target binary resembles the vulnerable version more than the patched one, it is flagged as unpatched. Another approach combines taint data-flow analysis and pattern matching to detect panic safety bugs, higher-order safety invariant violations, and concurrency issues \cite{bae2021rudra}. Rust’s HIR is analyzed for function declarations, trait implementations, and unsafe blocks, while MIR enables \textit{taint tracking} to detect lifetime violations and unsafe function assumptions. It also flags incorrect Send/Sync implementations by analyzing type definitions and API signatures for non-thread-safe methods. YUGA \cite{nitin2024uga} detects UAF and data races caused by incorrect lifetime annotations in Rust. It converts source code into HIR to extract structure definitions, function signatures, and lifetime constraints, infers borrow times, and uses pattern-based detection to flag lifetime inconsistencies. \textit{Flow- and field-sensitive intra-procedural alias analysis} on MIR confirms whether flagged value pairs alias the same memory location. \textbf{Machine learning:} it detects various bugs by learning context information from individual statements to entire programs. At the statement level, it models execution dependencies to identify vulnerabilities. One approach extracts control- and data-flow dependencies, structuring statement embeddings into a graph where nodes represent statements and edges capture dependencies \cite{hin2022linevd}. A \textit{graph attention network} learns topological structures, while an MLP classifier classifies vulnerable statements. At the function level, ML captures syntax, execution dependencies, structural dependencies, and historical bug indicators. Some models use syntax information, execution dependencies, and structural dependencies to detect vulnerabilities \cite{li2021vulnerability,liu2020cd,wang2023deepvd}. They extract features from statements, variables, types, function calls, AST paths, and control/data dependencies, transforming them into embeddings via tokenization. Sequence models like LSTMs and GRUs process embeddings, while ASTs, post-dominator trees, and exception flow graphs are embedded using tree-based (Tree-LSTM) or graph-based models (GNNs). These embeddings train classifiers such as \textit{GNNs, KNNs, or MLPs} for vulnerability prediction. Other models incorporate historical bug indicators, including function complexity metrics and repository metadata, to train classifiers or compute vulnerability scores for function-level detection \cite{meng2021bran, du2019leopard}. At the program level, one approach applies static analysis to extend or simplify structural graphs (e.g., ASTs, CGs, CFGs, DFGs, or PDGs), constructing extended flow graphs used as \textit{GNN} inputs for vulnerability detection at scale \cite{wen2023vulnerability,cheng2021deepwukong,wu2023learning}. Another approach leverages \textit{data mining} to identify frequent patterns and derive positive/negative rules for vulnerability detection. It extracts syntax features (e.g., variables, expressions) and semantic information (e.g., vulnerable code slices, control flow, function calls, condition checks) \cite{liang2016antminer,bian2018nar}. These features are analyzed to detect recurring vulnerability patterns and derive association rules.

\subsection{Conclusion of Bug Detection at IR Level}\vspace{-0.2em}
\begin{table*}[h]
    \centering
    \caption{Comparison of Bug Detection Techniques for C/C++ \& Rust}
    \vspace{-0.3em}
    \resizebox{\linewidth}{!}{  % Resizes table to fit within the page width
    \begin{tabular}{|c|c|c|c|c|}
        \hline
    \textbf{Paper} & \textbf{Techniques} & \textbf{ Bugs}&\textbf{Applications}&\textbf{Accuracy\&Overhead} \\
        \hline
     \cite{gabel2010scalable},\cite{kang2022tracer}&static &C/C++, logic &private,public &-,-\\
    \hline
\cite{wang2021amchex},\cite{wang2018check}&static &C/C++, missing-check , &public&46.2\% TP,-\\
\hline
\cite{yamaguchi2013chucky}&ML&C/C++, missing-check &public &-\\
\hline
\cite{deligiannis2015fast},\cite{bai2020effective},\cite{tan2011acomment},\cite{lepers2023ofence}&static &C/C++, concurrency &partially public, public&-,8.8\% FP,75\%,-\\
\hline
\cite{guo2024reorder},\cite{kasikci2017lazy}&hybrid &C/C++, concurrency &private,public &-,100\%\& 0.97\%\\
\hline
  \cite{UBITect},\cite{zhou2022non},\cite{guo2024precise}&static &C/C++, memory safety &public&59\% FP,-,37\%\\
\hline
  \cite{rigger2018sulong}&dynamic &C/C++, memory safety &public &-\\
  \hline
 \cite{yagemann2021automated}&hybrid &C/C++, memory safety &partially public, public & -\\
  \hline
  \cite{cao2022mvd}&ML&C/C++, memory safety &public &56.7\% \\
  \hline \cite{cui2023safedrop},\cite{hua2021rupair}&static &Rust, memory safety &private &-,48.3\%\\
\hline
\cite{mao2016rid},\cite{yang2022making}&static &C/C++,resource management &public &-,-\\
  \hline
   \cite{xu2020patch}& static &C/C++,unlisted&public &93.31\% \\

  \hline
\cite{hin2022linevd},\cite{li2021vulnerability},\cite{liu2020cd},\cite{wang2023deepvd},\cite{meng2021bran},\cite{wen2023vulnerability}, \cite{cheng2021deepwukong},\cite{wu2023learning},\cite{liang2016antminer}, \cite{bian2018nar}&ML&C/C++, unlisted&public &36\% F1,17.6\% F1,80.9\% F1,28.9\% recall,-,66.94\% F1,97.4\%,92.7\% F1,-,-\\
  \hline
   \cite{bae2021rudra}&static &Rust, logic, memory safety, concurrency &public &-\\
  \hline
   \cite{nitin2024uga}&static &Rust, memory safety, concurrency &public &87.5\% precision\\
  \hline
    \end{tabular}
    }
    \label{tab:cir}
\end{table*}

Table \ref{tab:cir} indicates that static analysis is the most widely used approach for detecting logic, memory safety, missing-check, concurrency, and resource management bugs in C/C++ and Rust at the IR level. Techniques rely on CGs,  CFGs and DFGs, and flow-sensitive, field-sensitive, alias, intra/inter-procedural, taint and data-flow analysis along with symbolic execution. Notable tools like UbiTect \cite{UBITect}, AMCHEX \cite{wang2021amchex}, and RUDRA \cite{bae2021rudra} effectively detect bugs such as use-before-initialization, missing security checks, and unsafe Rust patterns, with precision rates up to 87.5\%. Hybrid techniques like Reorder \cite{guo2024reorder} and FCatch \cite{kasikci2017lazy} combine static and runtime analysis to uncover concurrency and order violations with higher accuracy. Dynamic analysis, e.g., Safe Sulong \cite{rigger2018sulong}, captures runtime memory issues like UAFs via execution tracing. Machine learning is increasingly used for scalable detection. Models such as LineVD \cite{hin2022linevd}, DeepWukong \cite{cheng2021deepwukong}, and YUGA \cite{nitin2024uga} apply GNNs or LSTM-based architectures to learn vulnerability patterns from control/data dependencies, achieving high F1 scores (up to 97.4\%). However, the rest of solutions, except for ML-based techniques, do not report accuracy or overhead metrics. Instead, they typically present only the number of bugs detected.

\section{C/C++ \& Rust Bug Detection at Source Code Level}\label{sec:csource}

\subsection{Logic Bugs} Logic bugs at the source code level, including fast-path errors, anomalous variable-constant pairings, and protocol evolution inconsistencies, are detected using static analysis and machine learning. \textbf{Static analysis:} it identifies logic bugs through specification checking and symbolic execution. PALLAS \cite{huang2017pallas} detects fast-path logic errors by merging fast-path code and headers, constructing a control-flow graph, and applying symbolic execution with user-specified annotations (e.g., @immutable for immutable variables, @cond for necessary condition checks, and @order for proper conditional evaluation ordering). It also detects mismatched fast-path and slow-path return values, missing fault-handling, and uncoordinated updates between path states and associated data structures. \textbf{Machine learning:} it enhances static analysis through data mining \cite{lawall2010automated} and NLP \cite{chen2023ebugdec}.  \textit{Data-mining-assisted} static analysis detects variable-constant pairing bugs by constructing a bipartite graph linking constants to variables and clustering constants with similar usage profiles from flow-sensitive analysis. Clustering is performed using Euclidean and Jaccard similarity metrics, and anomalies are ranked by suspiciousness metrics such as cluster size and association strength, classifying them as name bugs, value bugs, or context bugs. \textit{NLP-assisted} static analysis detects RFC-evolutionary bugs by constructing evolutionary trees from RFC documents, extracting packet field meta-information (e.g., name, offset, size), and reconstructing packet hierarchies.It applies a predominator-based rule violation algorithm to detect implementation failures that do not adapt to new RFC conditions or lack necessary error handling.

\subsection{Multiple types of bugs } 
Machine learning detects source code-level bugs at the function and program levels. Function-level detection identifies vulnerable functions by analyzing high-level structures and execution contexts.  TROVON \cite{garg2022learning} focuses on pattern-based vulnerability detection by comparing vulnerable code fragments with their security fixes. It abstracts functions by replacing user-defined names and comments with generic identifiers, then processes vulnerable-fixed function pairs using an \textit{LSTM} model to detect potential bugs based on code differences. Program-level detection identifies bugs by analyzing entire programs using graph-based machine learning, inter-procedural slicing, and representation learning.  VCCFinder \cite{perl2015vccfinder} detects vulnerability-contributing commits by extracting repository-, author-, commit-, and function-level features from version control systems, classifying them using an \textit{SVM} trained on a Bag-of-Words model.   LSTMF \cite{lin2019software} applies deep-learning-based representation learning with \textit{Bi-LSTM} networks to learn vulnerable programming patterns. It converts code into vector embeddings using Word2Vec with a Continuous Bag-of-Words model, processes them with two \textit{Bi-LSTM} networks trained on different datasets, and trains a \textit{random forest} classifier. LSTMF also applies transfer learning to adapt to new projects and predict vulnerability likelihood.

\subsection{Conclusion of Bug Detection at Source Code Level}
%\begin{table*}[h]
%    \centering
%    \caption{Comparison of Bug Detection Techniques for C/C++}
    
%    \resizebox{\linewidth}{!}{  % Resizes table to fit within the page width
%    \begin{tabular}{|c|c|c|c|c|}
 %       \hline
 %       \textbf{Paper} & \textbf{Analysis Techniques} & \textbf{ Bugs}&Applications&Accuracy\&Overhead \\
 %       \hline
 %       \cite{huang2017pallas}& static  &C/C++, logic bugs&partially public    &69\%  \\
 %       \hline
        %\cite{lawall2010automated},\cite{chen2023ebugdec}& ML &C/C++, logic bugs& public   &70\%, 37.3s\\
 %       \hline \cite{garg2022learning},\cite{perl2015vccfinder},\cite{lin2019software}& ML &C/C++,unlisted & public  &85\% recall,99\% recall,56\%\\
 %       \hline
%    \end{tabular}
%    }
%    \label{tab:csource}
%\end{table*}
%precision, efficiency, techniques, benchmarks, bug types,
Static analysis methods like PALLAS \cite{huang2017pallas} employ symbolic execution and user-annotated specifications on source code to identify errors in fast-path code logic. It reports 69\% detection accuracy. Machine learning-assisted analysis improves detection by identifying structural and semantic patterns within the source code. For example, data-mining approaches \cite{lawall2010automated} use clustering and statistical similarity metrics to detect variable-constant mismatch bugs, while NLP-based techniques \cite{chen2023ebugdec} extract and analyze protocol specifications to find RFC-evolution inconsistencies. These methods offer strong performance, with accuracies around 70\% and manageable computational overhead (e.g., 37.3 seconds).

\section{Java \&Android IR-Level Bug Detection}\label{sec:javair}
\subsection{Logic Bugs}   
Logic bugs at the IR level, including functional inconsistencies, API misuses, incorrect condition handling and semantic errors, are typically detected using static, dynamic, hybrid analysis and machine learning. \textbf{Static analysis:} it detects logic bugs by constructing ASTs, CFGs and DFGs and applying data-flow analysis, flow-sensitive, context-sensitive, field-sensitive, and intra- and inter-procedural analysis.  Seader \cite{zhang2022example} detects and repairs security API misuses in Java libraries by analyzing secure and insecure code pairs, extracting ASTs, and detecting statement- and expression-level edits. It derives vulnerability-repair patterns using \textit{intra-procedural data flow analysis} and stores security APIs with abstract fixes as JSON templates. \textit{Inter-procedural backward slicing} matches WALA bytecode against stored patterns, automatically generating fixes for security API misuses. CryptoGuard \cite{rahaman2019cryptoguard} detects cryptographic and SSL/TLS API misuse bugs by applying \textit{flow-, context-, and field-sensitive data-flow analysis} combined with forward and backward program slicing. It constructs CGs, CFGs, and DFGs from Jimple, tracing inter-procedural dependencies. Backward slicing detects hardcoded secrets, weak encryption, and improper SSL/TLS validation, while forward slicing tracks insecure configuration propagation. Sensitive parameters (e.g., cryptographic keys, PRNG seeds, SSL configurations) are monitored across method calls to flag weak cryptographic keys, insecure random number generators, and missing certificate validation.  \textbf{Dynamic analysis:} it detects logic bugs by fuzzing. Genie \cite{su2021fully} detects functional bugs in Android apps using independent view \textit{fuzzing}. It builds a GUI transition model by generating random test seeds, exploring the app, and extracting GUI trees. Independent views, where interacting with one GUI element should not alter unrelated UI elements, are identified. Genie mutates interactions with inactive views and detects logic bugs by analyzing discrepancies between the original and mutated runs. Odin \cite{wang2022detecting} identifies non-crashing functional bugs by detecting deviant behaviors. It executes the target app with generated event sequences, captures GUI execution traces, and constructs a GUI model ensuring deterministic state transitions. Odin improves test coverage using a random walk simulation algorithm to prioritize under-explored GUI states and generate additional test inputs using \textit{fuzzing}. A hierarchical clustering algorithm groups observed GUI states, flagging unexpected small clusters as potential anomalies. 

\textbf{Hybrid analysis:} it detects logic bugs by pattern matching with dynamic validation. DCDroid \cite{wang2019dcdroid} detects SSL/TLS API misuse vulnerabilities in Android apps by generating call graphs from Smali code to trace SSL-handling functions (e.g., X509TrustManager, HostnameVerifier). Static analysis identifies entry points in Activities and Services where vulnerable code could execute. Automated UI interactions dynamically test execution paths. To confirm vulnerabilities, DCDroid uses mitmproxy to perform Man-in-the-Middle attacks, intercepting HTTPS traffic and using VPNService, UsageStatsManager, and PackageManager to correlate app behavior with network traffic. If HTTPS traffic observed on the mobile device matches intercepted mitmproxy traffic, the vulnerability is confirmed. \textbf{Machine learning:} it analyzes control/data dependencies and syntactic structures. CFGNN \cite{zhang2023detecting} applies a CFG-based \textit{GNN model} to detect condition-related bugs. It extracts CFGs from methods, tokenizes statements, and encodes them with a \textit{BiLSTM} to preserve sequential context. A graph-structured \textit{LSTM} captures long-range control-flow dependencies, while an API-aware attention mechanism highlights nodes containing API-related condition expressions. CFGNN classifies methods associated with condition-handling errors. Bugram \cite{wang2016bugram} detects semantic errors using n-gram language models. It parses source code into ASTs, converts them into token sequences, and trains \textit{n-gram models} (2 to 10 tokens in length) to learn probability distributions over token patterns. Token sequences with low probability scores across multiple models are flagged as potential semantic errors. LineFlowDP \cite{yang2024lineflowdp} enhances semantic error detection by applying \textit{GNNs} to PDGs. It categorizes PDG nodes into variable definitions, conditions, and execution nodes while classifying control-flow edges as True, False, and Next to  reflect execution semantics. To preserve execution order, it prioritizes control dependencies (True$>$Next$>$False) and extends code line semantics through backward and forward control/data flow dependencies. Doc2Vec encodes code lines into low-dimensional embeddings, which are processed by a \textit{Relational GNN} for file-level defect classification. Social network centrality ranks code lines by risk scores for defect localization.

\subsection{Performance Bugs} 
Performance bugs at the IR level include software hangs, algorithmic complexity issues,  performance cascading, lengthy operations, view holder violations, energy inefficiencies, and execution hotspots, detected through static, dynamic, hybrid analysis and machine learning. \textbf{Static analysis:} it detects performance bugs by constructing CGs,   CFGs and DFGs and applying intra/inter-procedural, flow sensitive, data-flow sensitive analysis with pattern matching. DScope \cite{dai2018dscope} detects software hangs caused by data corruption by decompiling bytecode into Jimple using Soot, constructing control and data-flow graphs, and identifying loops with exit conditions dependent on external data (e.g., I/O operations) using \textit{flow sensitive, data-flow sensitive analysis}. \textit{Loop-bound and stride analysis} filter out irrelevant loops. PerfChecker \cite{liu2014characterizing} identifies lengthy main-thread operations and view holder violations by analyzing lifecycle and event handler checkpoints, constructing call graphs, and detecting transitive calls to resource-intensive APIs such as networking and file I/O using \textit{intra/inter-procedural analysis}. It also verifies view reuse correctness in getView() callbacks using program dependency graphs. \textbf{Dynamic analysis:} it identifies performance issues through real-time execution monitoring \cite{jovic2011catch} or fuzzing to maximize resource consumption \cite{blair2022hotfuzz}.  \textit{Execution monitoring} analyzes method invocations, return values, and long-latency calls, profiling method executions and using selective call pruning to isolate performance bottlenecks. \textit{Fuzzing} detects algorithmic complexity vulnerabilities in Java libraries by treating each method in the target Java library as an independent entry point. It uses the Class Hierarchy Graph (CHG) for seed generation, instantiating concrete classes for abstract classes and interfaces. Random walks over CHG ensure diverse class selection. Genetic algorithms are then applied to mutate inputs to maximize execution time and memory consumption.  A combined approach integrates execution monitoring and fuzzing to detect energy   bugs and hotspots\cite{banerjee2014detecting}, building event flow graphs from user interactions, selecting energy-intensive traces, and \textit{monitoring} I/O system calls and power usage. By dynamically adjusting inputs, \textit{fuzzing} explores new traces, identifying system call sequences linked to abnormal power consumption and confirming energy bugs, while hotspots are flagged by comparing energy usage across traces. \textbf{Hybrid analysis:} it combines loop-bound analysis, concolic execution, runtime profiling, and fuzzing. detects performance cascading bugs in distributed systems \cite{li2018pcatch} by analyzing small-scale execution traces to designate critical execution areas as sinks, tracking execution dependencies and resource contention, and integrating static \textit{loop-bound analysis} with \textit{runtime profiling} to identify non-scalable code regions. Another approach detects algorithmic complexity vulnerabilities using \textit{fuzzing and concolic execution} \cite{noller2018badger}, generating inputs, constructing symbolic execution trees, identifying worst-case nodes with increasing code coverage or computational cost, and refining inputs iteratively by symbolic execution. \textbf{Machine learning} it detects energy bugs by modeling system operations and analyzing deviations in resource consumption patterns \cite{zhu2019evaluation}. A Lasso-based \textit{linear regression model} is trained on labeled datasets with system call features such as context switches, page faults, and file/socket operations. The trained model analyzes commit histories in updated software versions to detect abnormal energy spikes.

\subsection{Missing-check Bugs} 
Missing-check bugs at the IR level, including missing security checks, missing input validation, component hijacking bugs, permission re-delegation vulnerabilities,  Inter-Application Communication bugs, and missing-authorization, are detected through static, dynamic, hybrid analysis and machine learning.  \textbf{Static analysis} it detects missing-check bugs by constructing CGs, CFGs and DFGs, and applying  inter-procedural, context-, field-, flow- and  alias, data-flow analysis and symbolic execution.
MPChecker \cite{lu2022detecting} and WeChecker \cite{cui2015wechecker} identify missing security checks by analyzing unprotected privileged operations. MPChecker applies \textit{inter-procedural and field-sensitive data-flow analysis} to detect privileged operations and system-critical variables from logs, access patterns, and error conditions. It analyzes access patterns to user and system data and identifies permission-checking APIs using post-dominance relationships in control flows. It applies \textit{inter-procedural data-flow analysis} to detect unguarded privileged operations. Similarly, WeChecker constructs CFGs from Jimple IR, extracts components, intent filters, and callbacks from Manifest and XML files, and builds a CG. It marks unprotected components as entry points, analyzes the CG, and applies SUSI-based \textit{inter-procedural source-sink analysis} for capability leak detection. \textit{Alias-, flow-, context-, and field-sensitive analysis} identify sensitive data returned by permission-protected calls. ACO-Solver \cite{thome2017search} detects missing input sanitization checks by using \textit{symbolic execution} to solve supported attack constraints and Ant Colony Optimization to address unsupported constraints. ACO-Solver simulates artificial ants, each representing a candidate attack string, to explore paths likely to bypass sanitization. If a SAT solver confirms feasibility, the system flags the vulnerability. CHEX \cite{lu2012chex} and ARF \cite{gorski2019arf} detect component hijacking and permission re-delegation vulnerabilities, respectively. CHEX constructs CGs and DFGs on WALA IR, extracts entry points from Manifest and overridden framework methods, and applies \textit{context- and field-sensitive analysis} to generate split data-flow summaries (SDSs), These track intra-split flows, linking them across execution sequences to detect unprotected source-to-sink flow via \textit{inter-procedural analysis}. ARF identifies cases where a deputy entry point invokes a target entry point without enforcing necessary permission checks by constructing call graphs, extracting authorization checks, and identifying call paths between entry points accessible to third-party apps. \textbf{Dynamic analysis:} it identifies missing-check bugs using dynamic taint analysis, fuzzing and execution monitoring. Rivulet \cite{hough2020revealing} and IntentDroid \cite{hay2015dynamic} detect input validation and IAC vulnerabilities via \textit{taint analysis} and \textit{bytecode instrumentation}. Rivulet tracks user-controlled input propagation in Java bytecode using ASM instrumentation, flagging cases where untrusted data reaches sensitive sinks. It then generates attack payloads (e.g., XSS, SQL injection) and verifies whether sanitization checks are missing. IntentDroid instruments Dalvik bytecode, identifies IAC entry points from Manifest files, and executes crafted intents and attack payloads to assess enforcement of security checks, detecting potential user data leaks. PAIR \cite{liu2021privilege} and MoSSOT \cite{shi2019mossot} detect authorization bugs by modeling Remote Procedure Call (RPC) and SSO authorization mechanisms, respectively. PAIR analyzes Ant Group’s RPC system using cross-system tracing to construct behavioral dependency graphs (BDGs) capturing service dependencies. It decomposes RPCs into RPClets, prioritizing those handling sensitive data using entropy analysis of encrypted payloads. To detect authorization flaws, PAIR modifies high-risk RPClet requests by replacing user identifiers with test identifiers and replaying them. If the BDG remains unchanged, the system flags a bug. MoSSOT automates UI interactions to reach SSO login pages, captures network traces between mobile apps and identity provider servers, and applies differential analysis to infer authentication parameters. It generates HTTP(S) test cases, \textit{fuzzes} login credentials (e.g., access tokens, user IDs), and compares UI/network behaviors to detect unauthorized login attempts. AuthScope \cite{zuo2017authscope} builds on MoSSOT’s approach by substituting protocol fields (e.g., security tokens) with values of small Euclidean distances. Modified requests are sent to the server, and server responses are analyzed to identify authorization flaws. 

\textbf{Hybrid analysis:} it integrates static analysis with fuzzing for detecting missing-check bugs. AuthDroid \cite{wang2015vulnerability} detects authorization bugs using a five-party model (user, SP server, RP server, RP app, and SP app). It reverse engineers RP apps, applies static \textit{pattern matching} to extract participants, and uses \textit{dynamic traffic analysis} to monitor request/response interactions. Differential \textit{fuzzing} modifies parameters (e.g., app ID, access token), replays requests, and detects weak credential handling and token-user binding flaws. Demissie \cite{demissie2020security} detects second-order permission re-delegation vulnerabilities, where a malicious app exploits a privileged app to dispatch privileged Intents to system components. It extracts public Android components from the app’s Manifest, analyzes bytecode to identify entry points, and detects inter-component communication (ICC) method invocations (e.g., startActivity()) dispatching privileged Intents.  Constant propagation using privileged Action strings and \textit{data-flow analysis} collect additional sinks. A call graph maps paths from sources to sinks, which are verified via \textit{fuzzing}. \textbf{Machine learning:} it analyzes semantic and syntactic patterns for inconsistencies. One approach integrates \textit{NLP} with static analysis to detect missing security checks \cite{ouairy2020confiance}. It constructs a method-term matrix from extracted API calls, conditions, and loop keywords, applying document frequency weighting and \textit{Latent Semantic Analysis} for dimensionality reduction. Clustering groups similar functions, and inter-cluster comparisons identify missing checks. An inter-procedural control-flow graph verifies whether missing checks in a method are compensated by caller methods. Another approach leverages AST-based pattern recognition for multi-class classification of missing-check bugs \cite{al2022improving}. It extracts ASTs from methods, parses CVE records for labels, and uses depth-first search to generate AST n-grams. These n-grams are converted into a bag-of-words representation, labeled with CWE types or non-vulnerable tags, and analyzed using a \textit{random forest classifier} to assess security risks.

\subsection{Concurrency Bugs} Concurrency bugs at the IR level include time-of-fault errors and data races, detected through static and hybrid analysis. \textbf{Static analysis:} it detects concurrency bugs by applying context-sensitive alias analysis. nAdroid \cite{fu2018nadroid} identifies  UAF bugs in Android applications by modeling event-driven and multi-threaded concurrency. Given an APK package, it unifies event callbacks and posted callbacks as threads to preserve causal dependencies. It then detects UAF ordering violations using context-sensitive heap object naming and \textit{alias analysis} to track object lifetimes and identify access-after-free scenarios across Android-specific components such as Handlers, AsyncTasks, and Services. \textbf{Hybrid analysis} it combines flow-sensitive, data-flow-sensitive, intra-procedural, and inter-procedural analysis with enforced execution.  FCatch \cite{liu2018fcatch} detects time-of-fault bugs, which occur when component failures interfere with concurrent operations at critical moments. It instruments Java programs using WALA for function tracing and Javassist for inserting monitoring functions before heap and static variable accesses.  By observing both fault-free and faulty runs in distributed systems, FCatch tracks resource access and fault tolerance operations, detecting conflicting read-write operations on shared resources. \textit{Control and data dependency analysis} identifies unprotected conflicts, and \textit{enforced execution} manipulates timing to trigger bugs. DCatch \cite{liu2017dcatch} detects data concurrency bugs by analyzing correct execution traces and constructing a happens-before (HB) model to capture concurrency and communication patterns.  It tracks inter-node communication, intra-node asynchronous events, and multi-threaded synchronization to identify execution sequences vulnerable to data races and deadlocks. DCatch collects traces of memory accesses, HB operations, and synchronization events, then builds an HB graph to detect vulnerable access pairs. It applies static pruning at \textit{intra-procedural, inter-procedural, and inter-node levels} before re-executing the target system under controlled conditions to confirm bugs.

\subsection{Side Channel Vulnerabilities}  
Side-channel vulnerabilities, particularly resource usage leaks, are detected using dynamic and hybrid analysis. \textbf{Dynamic analysis:} it detects side channel vulnerabilities by using fuzzing and runtime analysis. DIFFUZZ \cite{nilizadeh2019diffuzz} uses resource-guided \textit{fuzzing} to identify inputs that maximize differences in resource consumption between secret-dependent execution paths. It iteratively executes a program with varying secret inputs while measuring execution costs (e.g., time, memory, response size). Mutation-based fuzzing generates new inputs that increase cost variations or improve program coverage, with high-difference inputs re-added to the fuzzing queue. The process continues until a significant, exploitable side-channel leak is detected, indicating that an attacker could infer secret data from execution cost variations. ProcHarvester \cite{spreitzer2018procharvester} detects procfs information leaks in Android by identifying and exploiting side-channel vulnerabilities. It operates in four phases: (1) Exploration scans procfs resources (e.g., /proc/meminfo, /proc/stat) to identify system files that change in response to events like app launches or keyboard inputs. (2) Profiling collects time-series data by logging fluctuations in these files while triggering events via ADB commands, MonkeyRunner scripts, or manual interactions. (3) Analysis applies dynamic time warping (DTW) to detect correlations between system resource variations and triggered events, even in noisy data.  (4) Attack exploits identified leaks to infer user actions, operating in single-resource mode (analyzing individual procfs files) or multi-resource mode (combining multiple sources for higher accuracy). \textbf{Hybrid analysis:} it detectsside-channel vulnerabilities by constructing call graphs and integrating pointer analysis, alias analysis, and runtime analysis. Themis \cite{chen2017precise} detects resource-based side-channel vulnerabilities by verifying $\varepsilon$-bounded non-interference in resource consumption. It analyzes Java bytecode using pointer and alias analysis on a call graph to track variables influencing secret-dependent operations, identifies hot spots where resource consumption depends on secret data, and instruments the code with consume statements to measure execution time and response size.  Themis evaluates whether these hot spots adhere to $\varepsilon$-bounded non-interference, ensuring that resource usage variations remain within an acceptable threshold. If deviations exceed the threshold, it indicates that secret-dependent execution differences could be exploited.

\subsection{Resource Management Bugs} 
Resource management bugs, including resource leaks and data loss, are detected through static and dynamic analysis. \textbf{Static analysis:} it identifies resource management bugs by constructing data-flow graphs and applying context-sensitive \textit{data-flow analysis} \cite{lu2022detecting}. It tracks resource usage violations caused by lifecycle variations in two scenarios: (1) Skip scenarios, where lifecycle methods like onStop are skipped, leading to memory leaks when resources allocated in onCreate are never released. (2) Swap scenarios, where changes in lifecycle execution order (e.g., onStop executing before onSaveInstanceState) cause data loss when edit operations occur without a corresponding save operation. \textbf{Dynamic analysis:} it detects data loss by \textit{injecting faults} and monitoring applications' behavior \cite{riganelli2020data}. It constructs a GUI model to track visited states, prioritizing unexplored states prone to data loss. To expose faults, it injects fault-revealing actions such as filling input fields and triggering stop-start events (e.g., screen rotations) to simulate interruptions. It compares screenshots before and after an interruption to detect visual discrepancies and analyzes GUI properties for unexpected changes. If an app fails to persist user input, resets UI elements unexpectedly, or modifies the interface inconsistently, it flags the issue as a potential data loss fault.

\subsection{Multiple Types of Bugs} 
Various bugs are simultaneously detected at the IR level through similarity comparison and multi-tier static analysis. One approach detects vulnerable third-party libraries by extracting control-flow and opcode features from Smali and Jimple IRs and matching them against a database of known vulnerabilities and CVE records \cite{zhan2021atvhunter}. It isolates library code from host applications, applies \textit{fuzzy hashing and edit distance comparisons} to determine library versions, and cross-references them with CVE records to generate vulnerability reports. Another approach integrates multiple static analysis techniques to detect and patch vulnerabilities across six categories: inter-component communication, storage, web, cryptographic, runtime-permission, and networking issues \cite{gajrani2020vulvet}. It constructs control-flow and data-flow graphs from Jimple IR and applies \textit{flow-sensitive, data-flow, taint, parameter, return-value, and API analysis} to detect security flaws. It also examines Manifest files to identify weak permission enforcement and multitasking issues. Beyond detection, it automatically generates patches by instrumenting Jimple code and modifying control flow, issuing warnings for unresolved vulnerabilities.

\subsection{Conclusion of Bug Detection at IR Level}
\begin{table*}[h]
    \centering
    \caption{Comparison of Bug Detection Techniques Across Java \& Android}
    \vspace{-0.5em}
    \resizebox{\linewidth}{!}{  % Resizes table to fit within the page width
    \begin{tabular}{|c|c|c|c|c|}
        \hline
        \textbf{Paper} & \textbf{ Techniques} & \textbf{ Bugs}&Applications&Accuracy\&Overhead \\
        \hline
         \cite{zhang2022example},\cite{rahaman2019cryptoguard}& static & Java, logic & public  & 95\% precision\&72\% recall,98.61\% precision\&12.7s\\
        \hline \cite{su2021fully},\cite{wang2022detecting}& dynamic  & Android logic & partially public,public  &40.9\% TP,-\\
        \hline
         \cite{wang2019dcdroid} & hybrid  &Android, logic  & private & 54.3\% TR\\
        \hline
        \cite{dai2018dscope}& static  &Java, performance & public & 53.2\% TR \\
        \hline
        \cite{liu2014characterizing}& static  &Android, performance  &public  & 54\% TR \\
        \hline 
\cite{jovic2011catch},\cite{blair2022hotfuzz}& dynamic  &Java, performance  & private &-,-  \\
        \hline
     \cite{li2018pcatch},\cite{noller2018badger}& hybrid  &Java, performance  &partially public,public  &-,-  \\
        \hline
  \cite{zhu2019evaluation}& ML &Android performance  &public  &94\%   \\
        \hline
     \cite{vala},\cite{cui2015wechecker}  & static  &Android, missing-check  &public  &87.5\%\&30s,96\% precision\&96\% recall\&30s  \\
        \hline
     \cite{thome2017search}&static &Java, missing-check &public &4.7\%-100\% \\
    \hline
 \cite{lu2012chex},\cite{gorski2019arf}  &static &Android, missing-check &private,public &81\% TR\&37s,-\\
    \hline
   \cite{hough2020revealing},\cite{liu2021privilege}  &dynamic &Java, missing-check &private, public &-,0 FN\\
    \hline
    \cite{hay2015dynamic},\cite{shi2019mossot},\cite{zuo2017authscope}&dynamic &Android, missing-check &private &92\% recall,-,-\\
    \hline
     \cite{wang2015vulnerability},\cite{demissie2020security}&hybrid &Android, missing-check &private,public &86.2\%,- \\
    \hline
    \cite{fu2018nadroid}&static &Android, concurrency &public &-\\
    \hline
    \cite{liu2018fcatch},\cite{liu2017dcatch}&hybrid &Java, concurrency &private, public &-,-\\
    \hline
     \cite{nilizadeh2019diffuzz}&dynamic &Java, side channel &public &-\\
    \hline
     \cite{spreitzer2018procharvester}&dynamic &Android, side channel &private &73\% \\
    \hline
     \cite{chen2017precise}&hybrid &Java, side channel &private &-\\
    \hline
    \cite{vala}&static &Android, resource management &public &-\\
    \hline
    \cite{riganelli2020data}&dynamic &Android, resource management &public &75\% \\
    \hline
   \cite{zhan2021atvhunter}&static &Java, unlisted&public &90.55\% precision\& 88.79\% recall\\
    \hline
    \cite{gajrani2020vulvet}&static &Android, logic, missing-check &private &95.23\% precision\&97.5\% F1\\
    \hline
 \cite{zhang2023detecting},\cite{wang2016bugram},\cite{yang2024lineflowdp}& ML &Java, logic bugs& public   & 46.2\% F1,71.2\% precision,37.6\%\\
        \hline
    \cite{ouairy2020confiance},\cite{al2022improving} & ML & Java, missing-check bugs&  partially public,public &33\% F1,75\% F1\\
        \hline
    \end{tabular}
    }
    \label{tab:javaIR}
\end{table*}

As summarized in Table \ref{tab:javaIR}, static analysis remains a widely used approach for detecting logic, missing-check, and performance bugs, leveraging control-flow graphs (CFGs), data-flow graphs (DFGs), alias analysis, and symbolic execution. Tools such as Seader \cite{zhang2022example} and CryptoGuard \cite{rahaman2019cryptoguard} identify API misuse and cryptographic vulnerabilities in Java libraries, achieving high precision (95-98.61\%) with reasonable computational overhead.MPChecker \cite{lu2022detecting} and WeChecker \cite{cui2015wechecker} effectively analyze missing security checks in Android applications, reporting 87.5\%-96\% precision with execution times around 30s. Dynamic analysis is primarily used for detecting runtime vulnerabilities, including functional logic errors, concurrency issues, and side-channel attacks. Tools such as Genie \cite{su2021fully} and Odin \cite{wang2022detecting} detect GUI-related logic bugs in Android apps, achieving 40.9\% true positive (TP) rates. For missing-check vulnerabilities, IntentDroid \cite{hay2015dynamic} and MoSSOT \cite{shi2019mossot} apply taint tracking and fuzzing, achieving 92\% recall. Dynamic execution monitoring tools, such as ProcHarvester \cite{spreitzer2018procharvester}, successfully identify side-channel vulnerabilities in Android with 73\% detection accuracy. Hybrid analysis, combining static and dynamic techniques, offers enhanced precision and validation mechanisms for detecting logic, missing-check, concurrency, and side-channel vulnerabilities. DCDroid \cite{wang2019dcdroid} detects SSL/TLS API misuses using a mix of static call graph construction and dynamic HTTPS traffic interception, achieving 54.3\% true recall (TR). AuthDroid \cite{wang2015vulnerability} and Demissie et al. \cite{demissie2020security} apply hybrid differential fuzzing and pattern matching to detect Android authentication flaws, achieving 86.2\% accuracy. ML-based approaches are increasingly being used for semantic-aware bug detection, particularly for logic and missing-check bugs. CFGNN \cite{zhang2023detecting} and LineFlowDP \cite{yang2024lineflowdp} utilize graph neural networks (GNNs) to analyze control/data dependencies and program structures, achieving 46.2\% F1-score and 71.2\% precision. For missing-check vulnerabilities, Confiance \cite{ouairy2020confiance} applies NLP-based semantic analysis, achieving 33-75\% F1-scores, depending on dataset granularity.

\section{JavaScript, PHP \& Python Bug Detection at IR Level}\label{sec:jsir}

\subsection{Logic Bugs } 
Logic bugs such as loop control issues, API misuse, type errors, attribute errors, and loose comparisons are detected through static, dynamic, and hybrid analysis.  \textbf{Static analysis:} it detects logic bugs using taint analysis, symbolic execution, and pattern comparison. Torpedo \cite{olivo2015detecting} identifies loop control flow bugs in PHP web applications by using \textit{taint analysis} and \textit{symbolic execution} to track tainted database attributes and detect loops dependent on tainted attributes. It performs backward \textit{symbolic execution} to generate attack vectors for verification.
SAFEWAPI \cite{bae2014safewapi} identifies API misuse in JavaScript by analyzing Web API function calls against specifications. It builds a CFG, a DOM tree, extracts \textit{Web API specifications} from HTML documents, traverses the CFG to identify API-related nodes, retrieves type information from the heap, and validates calls based on types, argument counts, and return values against specifications. \textbf{Dynamic analysis:} it detects logic bugs by runtime monitoring and pattern matching. AsyncG \cite{sun2019reasoning} detects asynchronous API misuse by running  Node.js applications and constructing an Async Graph (AG) representing event-driven execution.   AG nodes represent callback registration, execution, triggering, and object binding, while edges capture execution flow and object bindings. It applies patterns to detect scheduling bugs (e.g., recursive micro-tasks), emitter bugs (e.g., dead emits), and promise bugs (e.g., double resolve). \textbf{Hybrid analysis:} it integrates taint tracking, symbolic execution, and runtime validation. LChecker \cite{li2021lchecker} detects loose comparison bugs in PHP applications using static taint analysis and dynamic validation. It defines three conditions: untrusted input in comparisons (Cond1), implicit type conversion (Cond2), and inconsistent type behavior (Cond3). It constructs CFGs and CGs, applies \textit{taint analysis} to identify untrusted comparisons, and infers operand types. An enhanced PHP interpreter monitors loose comparisons at runtime, performing parallel strict type comparisons to detect inconsistencies. Xu \emph{et al.} \cite{xu2016python} detect type and attribute errors in Python by combining dynamic tracing with symbolic execution. The system executes test inputs, converts statements to Static Single Assignment form, and expands execution states via \textit{constraint solving}. Type errors arise from failed subtype assertions, while attribute errors occur when accessing nonexistent attributes due to inconsistent initialization. An SMT solver verifies constraint violations and reports bugs with triggering inputs.

\subsection{Missing-check bugs}
Missing-check bugs at the IR level are detected through static, dynamic analysis and machine learning. \textbf{Static analysis:} it includes alias, interprocedural, taint, data-flow, flow- sensitive analysis, and subgraph matching. CmrfScanner \cite{yang2022cross} detects Cross-MiniApp Request Forgery in JavaScript mini-apps using \textit{alias analysis}. It scans for cross-miniapp API calls (e.g., \texttt{navigateToMiniProgram}), converts relevant code into ASTs, and checks for missing \texttt{appID} verification. TChecker \cite{luo2022tchecker} and HiddenCPG \cite{wi2022hiddencpg} detect missing input validation in PHP applications.  TChecker performs \textit{context-sensitive interprocedural taint analysis}, constructing a call graph via backward \textit{data-flow analysis} to link call sites to targets and check if tainted variables reach sensitive operations without sanitization. HiddenCPG extracts vulnerable snippets from CVEs and GitHub, converts them into Code Property Graphs (CPGs) using Joern, and applies subgraph matching with the VF2 algorithm to detect unsanitized inputs. ObjLupAnsys \cite{li2021detecting} identifies prototype pollution in Node.js using \textit{flow-, context-, and branch-sensitive taint analysis}. It constructs an Object Property Graph from the AST to trace attacker-controlled inputs that may override system object properties, checking if built-in functions can be redefined. \textbf{Dynamic analysis:} it uses taint analysis, fuzzing, and runtime monitoring. SSRFuzz \cite{wang2024urls} targets Server-Side Request Forgery (SSRF) in PHP by analyzing sinks and URL schemes that process user-supplied URLs. It generates probe payloads to test these sinks and flags functions as SSRF sinks if a crafted URL triggers a server-side request. Dynamic taint inference hooks these sinks at runtime, while a web crawler identifies unsanitized input points. SSRFuzz then injects fuzzed payloads into these points, sends test HTTP requests, and detects SSRF vulnerabilities using six monitoring strategies: HTTP/DNS out-of-band tracking, port monitoring, HTTP response/log analysis, and file interaction tracking. Steinhauser \emph{et al.} \cite{steinhauser2020database} propose a graybox analyzer for detecting reflected context-sensitive, stored context-insensitive, and stored context-sensitive XSS flaws. It mutates HTTP request components (e.g., GET/POST parameters, cookies, headers) with XSS payloads and submits modified requests to observe whether injected values appear in responses (reflected XSS). For stored XSS, it injects payloads into both HTTP requests and database responses via a custom interception protocol, detecting whether the input is executed later. Context-sensitive analysis identifies failures in input transformation that lead to execution. \textbf{Machine learning:} it enhances static/dynamic analysis by filtering false positives. Kim \emph{et al.} \cite{kim2020finding} detect client-side business flow tampering by recording user interactions and identifying DOM elements tied to business logic. They generate automation scripts to replay navigation, monitor DOM mutations, collect function call traces, and construct business control flow graphs. A machine learning model prioritizes functions based on execution frequency and call stack position. Automated tampering tests modify JavaScript bytecode at runtime to bypass or alter function calls and branch outcomes. The results are clustered using screenshot similarity and Tree Edit Distance for HTML, verifying whether business logic (e.g., paywalls, ads) is bypassed. WAP \cite{medeiros2014automatic} detects missing input validation in PHP (e.g., SQL Injection, XSS, OS Command Injection) by combining \textit{static taint analysis} with machine learning. It converts source code into ASTs and trace control-flow paths from entry points (e.g., \$\_GET) to sensitive sinks (e.g., database queries), and extracts attributes related to string manipulation, input validation, and SQL query handling. \textit{Random Forest and Multi-Layer Perceptron} classifiers, trained on historical vulnerabilities, filter false positives.

\subsection{Concurrency bugs}
At the IR level, concurrency bugs such as data races and atomicity violations can be detected through \textbf{dynamic analysis}. It monitors execution traces, constructs happens-before (HB) graphs, and applies heuristics to identify conflicts. NRace \cite{chang2021race} detects data races by executing test cases while tracking lifecycle events, resource operations, and control flow. It builds an HB graph where nodes represent tasks and edges capture execution order based on program structure, event registration, promise resolution, and FIFO semantics. A race condition is flagged when two tasks perform conflicting operations on the same resource without an HB relationship. NodeAV \cite{chang2019detecting} targets atomicity violations in Node.js by instrumenting source code and collecting execution traces that capture events and read/write operations on shared resources. It constructs an HB graph consistent with Node.js’s event-driven model and identifies atomic event pairs expected to run without interference. Violations are detected by observing interleaved events that match four patterns: Read-Write-Read, Write-Read-Write, Read-Write-Write, and Write-Write-Read. Node.fz \cite{davis2017node} uncovers hidden concurrency bugs in Node.js by perturbing the execution order of asynchronous events. It operates in three phases: (1) Hooking and Interception to capture events (e.g., timers, I/O) into controllable queues; (2) Schedule Fuzzing to simulate rare interleavings via shuffling, delaying, and de-multiplexing; and (3) Execution and Monitoring to run perturbed schedules while detecting crashes, assertion failures, or incorrect outputs.

\subsection{Conclusion of Bug Detection at IR Level}\vspace{-0.8em}  % Reduces space between section title and table

\begin{table*}[h]
    \centering
    \caption{Comparison of Bug Detection Techniques Across JavaScript, PHP \& Python}
    \vspace{-0.5em}  % Reduces space between caption and table body
    \resizebox{\linewidth}{!}{  % Resizes table to fit within the page width
    \begin{tabular}{|c|c|c|c|c|}
        \hline
        \textbf{Paper} & \textbf{ Techniques} & \textbf{ Bugs}&Datasets&Accuracy\&Overhead \\
        \hline
         \cite{olivo2015detecting} & static & PHP, logic &  public  &58.3 TR\\
        \hline
          \cite{bae2014safewapi} & static  &JavaScript, logic  & private & 7.9\% FP\\
        \hline
      \cite{sun2019reasoning}& dynamic  &JavaScript, logic  &public  &-\\
        \hline
     \cite{li2021lchecker}& hybrid  & PHP, logic  & public &89\%-95\% , 68\% TP , 20 ms\\
        \hline
\cite{xu2016python} & hybrid  &Python, logic & public  &95.97\% \& 99.80\% accuracy\\
        \hline
    \cite{luo2022tchecker},\cite{wi2022hiddencpg}& static  &PHP, missing-check  &private, public  &46.1\% precision\&661m,-\\
        \hline
        \cite{wang2024urls}& dynamic  &PHP, missing-check &partially public  &46.1\% precision, 661m\\
        \hline
        \cite{medeiros2014automatic}&  ML &PHP, missing-check  &public &92.1\% accuracy, 92.5\% precision\\
        \hline\cite{yang2022cross},\cite{li2021detecting} & static  &JavaScript, missing-check &private  &95.97\% \& 99.80\%,-\\
        \hline 
    \cite{steinhauser2020database}& dynamic  & JavaScript, missing-check  &private &-\\
        \hline
        \cite{kim2020finding}&ML&JavaScript, missing-check&private&98.15\% precision\&85.01\% recall\\
        \hline
\cite{chang2021race},\cite{chang2019detecting},\cite{davis2017node}&  dynamic  & JavaScript, concurrency   &partially public, public &0.9s-17.0s, -,92.1\%\&92.5\% precision\\
        \hline
%         \cite{son2011saferphp}&  static  & PHP, logic,missing-check   &public &94.5\% precision\\
%        \hline
    \end{tabular}
    }
    \label{tab:otherir}
\end{table*}
As shown in Table \ref{tab:otherir}, 
static analysis remains the most prevalent technique for detecting logic bugs and missing-check bugs at the IR level, especially in PHP and JavaScript. Tools such as Torpedo \cite{olivo2015detecting}, SAFEWAPI \cite{bae2014safewapi}, and TChecker \cite{luo2022tchecker} leverage techniques like taint analysis, symbolic execution, and data-flow tracking to detect security vulnerabilities with varying levels of precision. Dynamic analysis is particularly effective in identifying runtime-dependent vulnerabilities, including concurrency issues and server-side request forgery. Tools such as SSRFuzz \cite{wang2024urls}, NodeAV \cite{chang2019detecting}, and NRace \cite{chang2021race} employ runtime monitoring, fuzzing, and schedule perturbation to uncover subtle bugs, such as race conditions and atomicity violations. While dynamic analysis typically achieves high detection precision, with Kim \emph{et al.} \cite{kim2020finding} reporting 98.15\% precision and 85.01\% recall, it can also introduce significant runtime overhead. For example, SSRFuzz \cite{wang2024urls} reports an execution time of 661 minutes. Hybrid analysis, which integrates static and dynamic approaches, offers a balanced trade-off between accuracy and efficiency. Tools such as LChecker \cite{li2021lchecker} and the method proposed by Xu \emph{et al.} \cite{xu2016python} apply taint analysis, symbolic execution, and runtime validation to detect logic bugs in PHP and Python, achieving accuracy rates as high as 99.8\%. Besides, many solutions do not provide the information about their dataset or give the application names without specific versions.

 %Hybrid analysis, which combines static and dynamic techniques, offers a more balanced approach by leveraging static code patterns and runtime execution insights. LChecker and Xu et al. demonstrate this approach in PHP and Python, achieving high accuracy (up to 99.8\%) while mitigating false positives from static analysis. This method is particularly effective for logic bug detection, as it integrates syntactic and runtime behavioral analysis.

\section{JavaScript, PHP \& Python Bug Detection at Source Code Level}\label{sec:jssource}

\subsection{Logic bugs } Logic bugs at the source code level include API misuse, cross-project compatibility, and name bugs, which are detected using static analysis and machine learning. \textbf{Static analysis:} it detects logic bugs via pattern matching and symbolic execution. WeDetector \cite{wang2022characterizing} detects WeChat Mini-Program bugs by analyzing JavaScript for three patterns: incorrect platform-dependent API usage, incomplete layout adaptation, and improper asynchronous return handling. It constructs ASTs from index.js and util.js to extract function definitions and calls.  It detects sensitive API misuse (e.g., \texttt{wx.makeBluetoothPair}), layout adaptation issues (e.g., improper use of \texttt{wx.getSystemInfo}), and improper handling of asynchronous calls (e.g., \texttt{wx.request}). NSBAp \cite{ma2020impact} detects Python cross-project bugs by analyzing how upstream bugs affect downstream modules. It constructs a dependence network mapping inter-project call relations and tracks version-specific dependencies. Intra-module analysis encodes function call paths, applying \textit{constraint solving} to assess whether a bug propagates across modules.  \textbf{Machine learning:} it detects logic bugs by learning from positive-negative code pairs. DeepBugs \cite{pradel2018deepbugs} targets name-related bugs in JavaScript by analyzing variable names, function arguments, and operators using semantic representations It trains a neural network on correct code and buggy samples generated by swapping parameters, altering operators, or misusing variables from correct code. Code is embedded using Word2Vec to preserve semantic relationships between identifiers and literals, and a feedforward neural network classifies code as correct or buggy.

%\subsection{Missing-check Bugs}
% \textbf{Static analysis} methods include signature comparison. 

\subsection{Multiple Types of Bugs}
Logic errors, concurrency issues, and missing checks can be simultaneously detected at the source code level using  multi-tier \textbf{static analysis} and pattern matching. RADAR \cite{10.1145/2814270.2814272} detects event-handling errors in Node.js by analyzing an event-based call graph composed of direct call edges, emit edges for event triggers, listen edges for listener registrations, and may-happen-before edges for execution order. It detects dead listeners when a listener is never triggered, dead emits when an event has no listeners, mismatched synchronous and asynchronous calls when improper mixing causes race conditions, and unreachable functions when they are never invoked or referenced.
VuDeFr \cite{kluban2022measuring} detects prototype pollution and Regular Expression Denial of Service vulnerabilities in JavaScript using custom Semgrep patterns to identify unsafe object property assignments and malicious regular expressions. It enhances efficiency with \textit{textual similarity search}, comparing function hashes against a vulnerability dataset and flagging functions that match predefined patterns or known vulnerability hashes. KlDvD \cite{kluban2024detecting} builds on VuDeFr by identifying and verifying these vulnerabilities. It uses Semgrep patterns and hashing to identify near-duplicate vulnerabilities, and verifies exploitability via \textit{static taint analysis}. File dependency graphs are constructed to track taint propagation from entry points to vulnerable functions.

\subsection{Conclusion of Bug Detection at Source Code Level}
Static analysis remains the dominant approach, detecting logic, concurrency, and missing-check bugs primarily in JavaScript and Python source code, with some works explicitly addressing multiple types of bugs. Only one work uses ML-based techniques to detect logic bugs in JavaScript, due to a lack of available buggy open source projects. Performance evaluation reveals that static analysis techniques report a 7.9\% false positive rate for Python logic bugs \cite{ma2020impact} and 58.3\% true positive rate for JavaScript logic bugs, while only reporting execution times for JavaScript multi-type bug detection  range from 0.9s to 17.0s. ML-based techniques, in contrast, achieve a 68\% true positive rate with a 20ms overhead, showing potential in reducing detection time.

\section{Cross Language IR-Level Bug Detection}\label{sec:crossir}

\subsection{Logic Bugs}
Logic bugs spanning multiple languages, such as mishandled JNI exceptions and compiler errors, can be detected using static analysis and machine learning. \textbf{Static analysis:} it employs interprocedural, data-flow, and taint analysis. SBSAf \cite{li2009finding} detects and recovers from mishandled JNI exceptions in Java Native Interface (JNI) programs, where exceptions in native code (C/C++) are not automatically handled by the JVM. Mishandling occurs when a pending exception is followed by an unsafe operation (e.g., strcpy). SBSAf tracks exception states via \textit{interprocedural data-flow analysis} and applies \textit{static taint analysis} on a pointer graph to detect unsafe operations near exception states. Unsafe operations not on a whitelist trigger warnings, and ExceptionClear statements are inserted to recover from mishandling. \textbf{Machine learning:} it enhances logic bug detection by generating diverse test cases.  ComFuzz \cite{ye2023generative} applies a \textit{Transformer-based neural model} to generate syntactically valid, semantically diverse test programs for Java and JavaScript compilers. It trains on GitHub repositories and buggy test suites, extracting ASTs, tokenizing function-level blocks using Byte Pair Encoding, and using MCMC sampling to prioritize test cases. Differential analysis detects parsing or optimization bugs by comparing compiled test outputs, while mutation-based iterative testing refines programs that cause crashes or incorrect outputs.

 \subsection{Memory safety bug }
Memory safety bugs,such as double free, use-after-free, and buffer overflow, can span multiple programming languages and are detectable through static analysis. \textbf{Static analysis:} it converts different programming languages into the same IR and applies inter-procedural flow analysis. ACORN \cite{xia2023acorn} detects memory safety bugs in multilingual Rust programs by using Wasm as a unified IR for cross-language analysis. It translates Rust and C components into Wasm using formalized rules that map types, memory operations, and control flow constructs. ACORN then builds CFGs and CGs  to model Rust-C interactions. It applies Wasmati and Wasabi to traverse CFGs \textit{interprocedurally} and analyze memory operations, detecting double-free, use-after-free, and buffer overflow. CRUST \cite{hu2022crust}  translates Rust and C into a common intermediate representation, CRUSTIR, which integrates control flow, data flow, type information, lifetime management, and ownership semantics, enabling precise memory safety analysis. It applies existing bug detection tools (rustc, Miri, MirChecker, and Rudra) to identify issues such as use-after-free, double-free, out-of-bounds access, stack overflow, and integer overflow.

%\subsection{Missing-check Bugs}
%Missing-check bugs detected across multiple languages include missing input validation and access control flaws, which can be identified through dynamic analysis by monitoring HTTP and SQL traces.  BATMAN \cite{li2014automated} simulates user interactions and captures HTTP and SQL traces to detect access control bugs. It infers role-level policies by ensuring lower-privileged roles cannot access higher-privileged data and derives user-level policies by linking SQL parameters to user identities. Role-based tests verify whether lower-privileged roles can execute restricted database operations, while user-based tests check if a user can manipulate database access parameters linked to another user’s identity. Detected violations are reported as access control flaws.

\subsection{Multiple Types of Bugs}
Various types of bugs across languages can be detected at the IR level using static analysis, dynamic analysis, and machine learning. \textbf{Static analysis:} it uses pattern matching and structural similarity. Pewny \emph{et al.} \cite{pewny2014leveraging} introduced tree edit distance-based equational matching to detect vulnerabilities in C/C++ and Objective C. It extracts disassembly and CFGs, summarizes basic blocks as expression trees, and matches them to known buggy code signatures using tree edit distance. Functions are ranked by similarity to reference signatures, highlighting likely vulnerable code regions. \textbf{Dynamic analysis:} it utilizes runtime monitoring and constraint validation. Jinn \cite{lee2010jinn} identifies foreign function interface (FFI) violations in JNI and Python/C programs using context-specific dynamic analysis based on state machine specifications. It models FFI constraints using state machines that define state transitions and language interactions, ensuring proper Java-Python-C integration. Jinn intercepts Java-C transitions, monitors runtime states, and validates FFI constraints by analyzing JNI function calls, exception state changes, resource management, and thread switches. \textbf{Machine learning:} it identifies bugs by analyzing historical data. Luo \emph{et al.} \cite{luo2022compact} propose Compact Abstract Graphs (CAGs) for vulnerability detection in Java and C. CAGs are derived from Abstract Graphs (AGs), built from ASTs with nodes representing tokens or syntactic properties and edges defining relationships. AGs are built by reversing AST edges, adding edges between sequential tokens, and linking terminal nodes to the root as a global sink. AGs are compressed by merging node sequences and aggregation structures, then embedded into a vector space for \textit{GNN-based} vulnerability classification.  Scaffle \cite{pradel2020scaffle} detects bugs in Android, IOS and PHP codebases by combining machine learning with an information retrieval-based search engine. Using historical crash-fix data, it trains a \textit{bi-directional RNN} at the line level to generate vector representations of words within each line and a \textit{bi-directional RNN} at the trace level to predict the relevance score of each line. The information retrieval-based search engine then queries the highest-ranking lines to retrieve  buggy files. 

\subsection{Conclusion of Bug Detection at IR Level}\vspace{-0.3em}  % Reduces space between section title and table
%\begin{table*}[h]
%    \centering
%    \caption{Comparison of Bug Detection Techniques for Cross-Languages}
%    \vspace{-0.5em}
%    \resizebox{\linewidth}{!}{  % Resizes table to fit within the page width
%    \begin{tabular}{|c|c|c|c|c|}
%        \hline
%        \textbf{Paper} & \textbf{ Techniques} & \textbf{ Bugs}&Dataset&Accuracy\&Overhead \\
%        \hline
 %        \cite{li2009finding} & static  & C/C++, Java, logic & public &15.4\% FP\\
%        \hline
%          \cite{ye2023generative} & static  &Java, JavaScript, logic  &public  &87.8\% TR \\
%        \hline
% \cite{xia2023acorn},\cite{hu2022crust}& static  & Rust, C, memory safety  & private &100\% precision\&89.04\% recall,84.1\%\\
%        \hline
%      \cite{pewny2014leveraging} & static  &C/C++, Objective C, missing checks,logic, memory safety  & private &40-60\% \\
%        \hline
%         \cite{lee2010jinn}& dynamic  & Python, C, logic, resource management, concurrency  &partially public &100\% accuracy, 14\% overhead\\
%        \hline
%       \cite{luo2022compact}&  ML & Java, C, unlisted&private &94.7\%-96.3\% \& 91.6\%-93.2\% \\
%        \hline
%         \cite{pradel2020scaffle}&  ML & Android, Objective C, PHP, unlisted  &public &40-60\% \\
%        \hline
%    \end{tabular}
%    }
%    \label{tab:crossir}
%\end{table*}
%precision, efficiency, techniques, benchmarks, bug types,
%Table \ref{tab:crossir} presents a comparison of bug detection techniques at the intermediate representation (IR) level, highlighting the cross-language applicability of different approaches. 

 Static techniques remain the predominant approach, particularly effective in cross-language scenarios involving Rust and C memory safety bugs, where they achieve high precision and recall (e.g., \cite{xia2023acorn}, \cite{hu2022crust}). In contrast, dynamic analysis offers enhanced coverage for concurrency and runtime resource management bugs, demonstrating strong accuracy with manageable overhead in cross-language settings, as exemplified by \cite{lee2010jinn}. ML-based approaches (e.g., \cite{luo2022compact}, \cite{pradel2020scaffle}) often operate on large-scale, real-world datasets, including industrial codebases, but typically do not explicitly delineate the specific bug types they detect. While these methods show potential for cross-language generalization, their detection performance tends to vary across different language and bug contexts.

%\section{Cross Language Source Code-Level Bug Detection} Various types of bugs across languages can be simultaneously identified at the source code level through static analysis and machine learning.

%\subsection{Multiple types of bugs }

%Static analysis identifies various bugs by vulnerable patterns. VDiOS \cite{reid2022extent} detects orphan vulnerabilities in reused code by analyzing file-level code reuse across open-source repositories. It traces commit history using World of Code (WoC) mappings to identify vulnerable file versions and their corresponding fixes, generating lists of vulnerable and patched blobs. VDiOS then searches WoC to find projects that have ever contained the vulnerable blobs, even if later modified or removed.  Using hosting platform APIs (e.g., GitHub API), it checks whether a project’s current version still contains a vulnerable blob (unpatched), a fixed blob (remediated), or an unknown modification with uncertain impact.

\section{Challenges, and Research Gaps}\label{sec:trends}
	\begin{figure*}[h]
		\centering
		\includegraphics[scale=0.31]{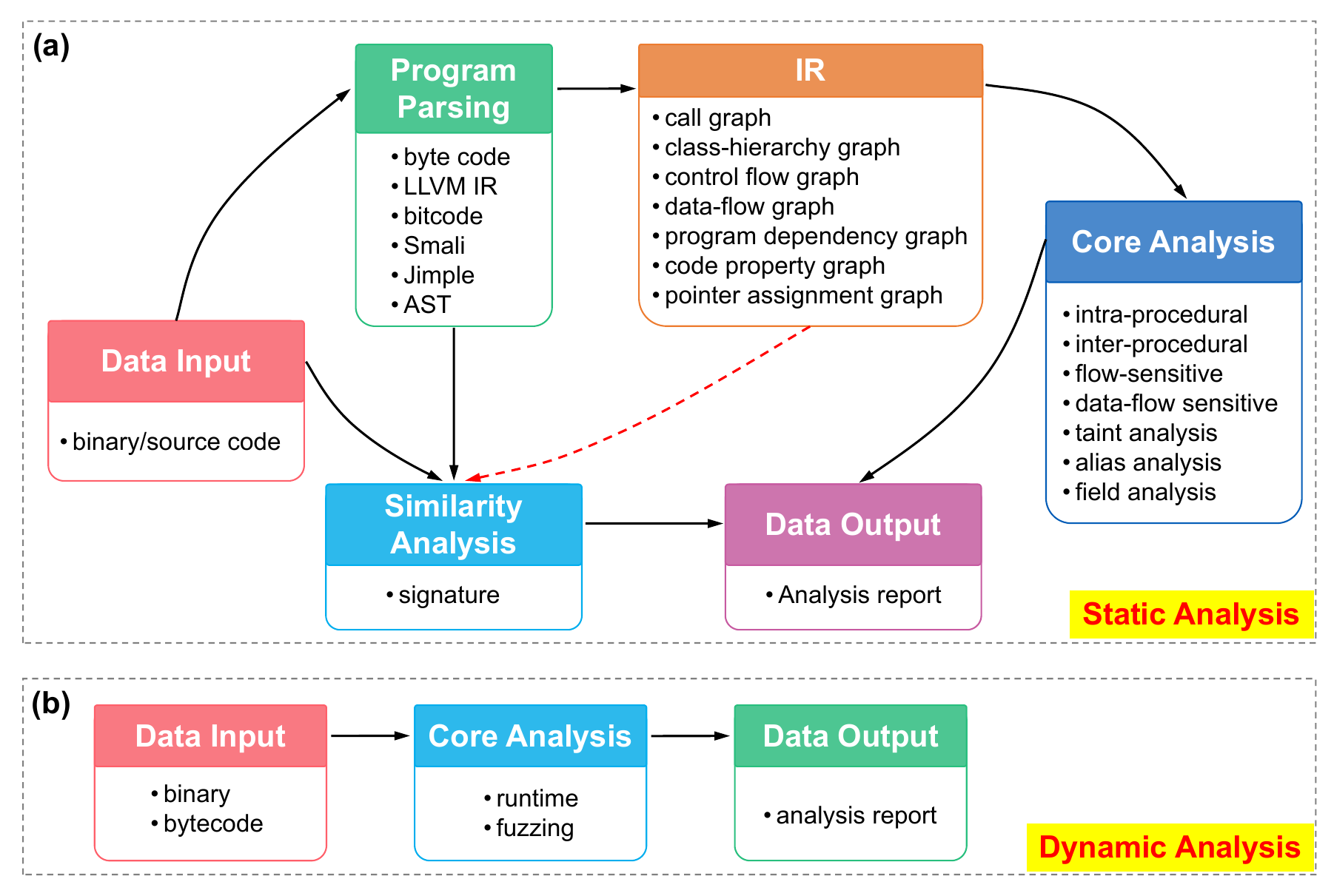}
		\caption{Overview of Analysis Techniques}
		\label{fig:analysis}
	\end{figure*}
Despite growing interest in software vulnerability detection, our survey identifies several critical research gaps. Java and Android, though widely used in enterprise and mobile applications, have received limited attention at the source code level, with only one notable work addressing this space \cite{sultana2023software} because of a limitation compounded by the scarcity of open-source applications for these languages. Performance bug detection remains largely limited to C/C++, Java, and Android, with few efforts targeting dynamically typed languages or cross-language performance issues, aside from \cite{xiao2015uncovering, tizpaz2020detecting} for JavaScript and Python. 

 Furthermore, at the IR level, research remains fragmented, with very few efforts targeting multiple bug types in languages like PHP, JavaScript, and Python—notably, only one work exists for PHP \cite{son2011saferphp}. Similarly, source code-level multi-bug detection in PHP and Python is severely underexplored. Among these, Python remains the least tested programming languages though it is the most popular language for machine learning applications. More generally, most ML-based techniques still struggle to learn semantically meaningful patterns from raw source code, limiting their pre-compilation bug detection capabilities across programming languages. Perhaps most notably, no current approach unifies multiple programming languages for source-level bug detection, revealing a fundamental gap in cross-language analysis at the source code level. This shortfall is particularly problematic as modern applications are inherently polyglot, often comprising components written in diverse languages that interact via APIs, system calls, or network protocols. In such systems, vulnerabilities originating in one component can propagate, posing system-wide security risks. Effective cross-language and cross-representation analysis is therefore not optional but essential.

 Finally, we observe notable concerns around reproducibility and transparency. Approximately 37.1\% of surveyed works fail to disclose dataset details or specify application versions, and only 45.7\% release their source code, making comparative evaluations difficult and hindering progress in the field. These findings collectively highlight an urgent need for comprehensive, reproducible, and language-agnostic vulnerability detection techniques, especially at the source code level, to support scalable, multi-bug, and cross-language analysis in real-world, heterogeneous software systems.

\section{Conclusion}\label{sec:conclusion}
The main analysis pipelines for vulnerability detection across programming languages are illustrated in Fig.\ref{fig:analysis}. Static analysis (a) often relies on intermediate representation languages, such as LLVM IR, or AST, to construct structural, control, or data dependencies (e.g., call graphs), and apply static analysis algorithms to identify potential vulnerabilities. Alternatively, some approaches directly extract meta-information from binaries, bytecode, or source code to generate signatures, which are then matched against known vulnerable patterns. Machine learning (red dashed line)  uses features derived from various IRs to train models for bug detection. Dynamic analysis (b) relies on runtime monitoring to capture buggy execution traces or uses fuzzing to trigger crashes. Hybrid analysis integrates static and dynamic techniques to enhance detection capability. 
Modern software systems are written in diverse languages, such as C/C++ for low-level system modules, and Java for enterprise applications. Each language introduces unique syntax, semantics, and vulnerability surfaces. Consequently, purely language-agnostic techniques may fail to capture language-specific bug patterns and nuances in program behavior. Furthermore, applications may be analyzed at different levels, source code, IR, or binary, depending on the development or deployment context, and each representation supports different forms of analysis. Effective vulnerability detection must therefore align with both the programming language characteristics and the analysis granularity. This survey underscores the pressing need for more robust, general-purpose tools and benchmarks that are capable of analyzing heterogeneous codebases across programming languages, while also adapting to the specific requirements of different bug types and representations.

%As modern applications are increasingly polyglot, composed of multiple components written in different languages, vulnerabilities in one component can propagate and compromise the entire system. Cross-language and cross-representation analysis is therefore essential for uncovering system-wide security risks. This survey highlights the need for the development of more robust, general-purpose vulnerability detection tools and benchmarks that span languages and abstractions. It also points to critical gaps in current methods, where new techniques are needed to address evolving challenges in either single or multi-language software ecosystems.

\bibliographystyle{ACM-Reference-Format}
\bibliography{sample-base}

\end{document}